%% file: Draft2d.tex
\documentclass[11pt]{article}
\pdfoutput=1

\usepackage{a4wide}
\usepackage{Packages}
\usepackage{Definitions}

\usepackage{setspace}

\hypersetup{pdfauthor={Robbert Dijkgraaf, Domenico Orlando, Susanne Reffert},pdftitle={Quantum Crystals and Spin Chains}}

\author{
  \begin{minipage}{.97\linewidth}
    \vspace{1cm}
    \begin{center}
      \begin{small}
        \textbf{Robbert Dijkgraaf}${}^{1,2}$, \textbf{Domenico Orlando}${}^3$ and \textbf{Susanne Reffert}${}^{2}$
      \end{small}
    \end{center}
    \vspace{1cm}
    \hspace{2cm}\begin{minipage}{.7\linewidth}
     {\it \begin{footnotesize}
     \begin{itemize}
	\item[${}^1$] KdV Institute for Mathematics, University of Amsterdam,\\
          Plantage Muidergracht 24, 1018 TV Amsterdam, The Netherlands.
	\item[${}^2$] Institute for Theoretical Physics, University of Amsterdam,\\
          Valckenierstraat 65, 1018 XE Amsterdam, The Netherlands. 
          \item[${}^3$] Institut de Physique, Universit\'e de Neuch\^atel,\\ 
	Rue Breguet 1, CH-2000 Neuch\^atel, Switzerland.
       \end{itemize}
     \end{footnotesize}}
    \end{minipage}
    \vspace{1cm}
  \end{minipage}
}

\date{}

\title{\vspace{1.5cm}
  \begin{huge}
    \textbf{Quantum Crystals and Spin Chains}
  \end{huge}
}

\begin{document}

\numberwithin{equation}{section}

\begin{titlepage}
  \maketitle
  \thispagestyle{empty}

  \vspace{-12cm}
  \begin{flushright}
    ITFA-2008-05
  \end{flushright}

  \vspace{14cm}

  \begin{center}
    \textsc{Abstract}\\
  \end{center}

  In this note, we discuss the quantum version of the melting crystal
  corner in one, two, and three dimensions, generalizing the treatment
  for the quantum dimer model. Using a mapping to spin chains we find
  that the two--dimensional case (growth of random partitions) is
  integrable and leads directly to the Hamiltonian of the Heisenberg
  \textsc{xxz} ferromagnet. The three--dimensional case of the melting
  crystal corner is described in terms of a system of coupled
  \textsc{xxz} spin chains. We give a conjecture for its mass gap and
  analyze the system numerically.

\end{titlepage}

\onehalfspace

\tableofcontents

\newpage

\input{Introduction}

\input{QuantumHamiltonian}

\input{OneDimension}

\input{FermionFormalism2d}

\input{ThreeDimensions}

\input{Conclusions}

\subsection*{Acknowledgements}

We would like to thank Paul Fendley for pointing us to the literature
on the quantum dimer. It is a pleasure to thank Luis Alvarez--Gaum\'e,
Jean--Sebastien Caux, Philippe Di~Francesco, Luciano Girardello, and
Bernard Nienhuis for illuminating discussions and Shannon Starr for
useful correspondence. Furthermore we would like to thank Nicolai
Reshetikhin for careful reading of the manuscript. D.O. would like to
thank the University of Amsterdam for hospitality. S.R. would like to
thank the University of Neuch\^atel for hospitality. The authors would
like to thank the V Simons Workshop in Stony Brook for hospitality,
where this work was initiated. The research of R.D. was supported by a
NWO Spinoza grant and the FOM program "String Theory and Quantum
Gravity."  D.O. is supported by the Swiss National Science Foundation
and by the EU under contract MRTN-CT-2004-005104.  S.R. is supported
by the EC's Marie Curie Research Training Network under the contract
\textsc{MRTN}-\textsc{CT}-2004-512194 "Superstrings".

\newpage
\appendix

\input{NPointFunctions}

\bibliography{DimerReferences}

\end{document}

%% file: Introduction.tex
\section{Introduction}

In this note, we study the quantization of the statistical system of a
melting crystal in one, two, and three dimensions.  In two and three
dimensions, the melting crystal configurations correspond to the
diagrams of random partitions, resp. plane partitions. Random
partitions appear in many branches of mathematics and physics, such as
Gromov--Witten theory, random matrix theory, Seiberg--Witten theory
and others, see \emph{e.g.}~\cite{okounkov3} for an overview. Also in
string theory, random partitions are of importance.
In~\cite{Okounkov:2003sp, Iqbal:2003ds}, the partition function of the
melting crystal corner given by the MacMahon function was shown to
equal the partition function of the topological string A--model with
target space $X=\mathbb{C}^3$.

In the spirit of stochastic quantization, we construct a quantum
Hamiltonian whose ground state corresponds to the classical steady
state distribution. The treatment is reminiscent of the work of
Rokhsar and Kivelson on quantum dimer models~\cite{Rokhsar1988} (the
perfect matchings of the dimer model on the hexagonal lattice being in
one--to--one correspondence with the configurations of the 3d crystal)
and Henley's study of the classical and quantum dynamics for systems
satisfying the detailed balance condition~\cite{Henley2003}. While the
quantum dimer model only keeps track of the total number of
configurations, the number of boxes of each crystal melting
configuration is taken into account in our system. In the partition
function, each configuration is weighted by $q^{\#\text{boxes}}$. The
quantum dimer model at the Rokhsar--Kivelson point corresponds
therefore to the limit $q\to 1$ of the melting quantum crystal.

The problem of the two--dimensional quantum crystal can be
reformulated in terms of a one dimensional fermionic system. The
system translates to an asymmetric, hard--core diffusion problem on a
half--full infinite chain. In fact, the quantum Hamiltonian we derive
is easily transformed into the Hamiltonian of the \textsc{xxz}
ferromagnetic Heisenberg spin chain. The ground state of the partition
problem corresponds to a particular sector (the half--filled) of the
kink ground state of the infinite chain. Using the technology of basic
hypergeometric series, we derive the limit
shape~\cite{Vershik:1996aa}, which corresponds to the integral of the
magnetization in the limit $q\to1$. Also the general expression for
$n$--point correlation functions is derived.

Based on our findings in the two--dimensional case, we can express the
3d system as a system of interlaced spin chains. The quantum crystal
corner is a rare example of a three--dimensional system which is
nonetheless fairly tractable. This is of course due to its very
constrained nature. Expressed in terms of spin chains, it becomes
manifest that its dynamics is not even fully two--dimensional.

Based on numerical evidence, we conjecture the mass gap of the 3d
quantum crystal to be the same as in the 1d and 2d systems.  More
generally, all these systems can be interpreted as directed random
walks on the graph of configurations, where going forward and going
back has different weights. Depending on the dimensionality of the
problem, the graph of configurations changes, but it always remains a
directed acyclic graph. This more general framework unifies the three
problems considered here and might also account for common properties,
such as the value of the mass gap.

The plan of this note is as follows. In
Section~\ref{sec:class-vs.-quant}, we explain the general procedure of
deriving the quantum Hamiltonian for a melting crystal. As a warm--up
we specialize in Sec.~\ref{sec:one-dimens-young} to the simplest
possible problem, namely 1d crystal melting. The two--dimensional
system (\ie\ random partitions) is discussed in
Section~\ref{sec:2d_partitions}, where the Hamiltonian is expressed in
terms of fermionic operators (Sec.~\ref{sec:fermion2d}) and mapped to
the \textsc{xxz} spin chain (Sec.~\ref{sec:XXZ}). The 1--point,
2--point, and n--point correlation functions are derived in
Sections~\ref{sec:one-point-function} to~\ref{sec:n-point-function},
while Sec.~\ref{sec:numerics2d} gives some numeric results for higher
eigenstates. In Section~\ref{sec:3d-hamiltonian}, the
three--dimensional system is discussed and numeric results for the
higher eigenstates of the 3d case are presented.  The conjecture for
the mass gap is stated in Sec.~\ref{sec:mass-gap}.

In Section~\ref{sec:conclusions}, we end with concluding remarks and
an outlook on the open problems.


%% file: QuantumHamiltonian.tex
\section{The quantum Hamiltonian}
\label{sec:class-vs.-quant}

In this section, we will explain the general framework of quantization
we will be using. The quantum dimer model of Rokhsar and
Kivelson~\cite{Rokhsar1988} is an example thereof. In fact, it is a
limit of the more general problem of quantum crystal melting.

\subsection{The general framework}
\label{sec:more-general-sense}

Consider a classical system with a discrete set of configurations
$\set{C_i}$, such as a lattice model. Let $\bar \HH$ be the energy
functional that associates to each configuration its energy $\bar \HH
(C_i)$. If we consider the canonical ensemble, the probability of
finding the configuration $C_i$ is given by its Boltzmann weight
\begin{equation}
  p (C_i) = \frac{e^{-\beta \bar \HH (C_i)}}{Z} \, ,  
\end{equation}
where $\beta$ is the inverse temperature and $Z$ is the partition function
\begin{equation}
  Z = \sum_i e^{-\beta \bar \HH (C_i)} \, .
\end{equation}

We want to quantize this system. To do so we introduce the separable
Hilbert space $\mathscr{H}$ spanned by the orthonormal vectors
$\ket{C_i}$ which are in one--to--one correspondence with the
classical configurations. Each vector in $\ket{\psi} \in \mathscr{H}$
defines a probability measure on the $C_i$ as follows:
\begin{equation}\label{eq:measure}
  m_{\psi} (C_i) = \frac{\abs{ \braket{ \psi | C_i} }^2}{\norm{\psi}^2} \, ,
\end{equation}
where the scalar product is understood in $l^2$.

Consider the state $\ket{\Psi_0}$ defined by
\begin{equation}
  \ket{\Psi_0} = \sum_i e^{- \beta \bar \HH(C_i) /2} \ket{C_i} \, .  
\end{equation}
It is immediate to verify that the measure associated to $\ket{\Psi_0}$
is precisely the Boltzmann distribution for the canonical ensemble, up
to the normalization fixed by
\begin{equation}
  \label{eq:Psi0-Norm}
  \braket{ \Psi_0 | \Psi_0 } = Z \, .
\end{equation}
To be more explicit, consider a physical quantity $Q$ that can be
measured on the classical configurations $\set{C_i}$. On the canonical
ensemble, the statistical average of $Q$ is given by
\begin{equation}
  \braket{Q}_{\text{cl}} = \frac{1}{Z} \sum_i Q (C_i) e^{-\beta \bar \HH (C_i)} \, .  
\end{equation}
To such a quantity we can associate an operator $\hat Q$ in the Hilbert
space $\mathscr{H}$ which is diagonal in the basis given by the $\ket{C_i}$,
\begin{equation}
  \hat Q \ket{C_i} = Q(C_i) \ket{C_i} \, .  
\end{equation}
For any such $Q$, the statistical average thus coincides with the
expectation value of $\hat Q$ on the state $\ket{\Psi_0}$ (which is
equivalent to the above statement of the measure corresponding to the
Boltzmann distribution):
\begin{equation}
  \frac{\braket{\Psi_0 | \hat Q | \Psi_0}}{\braket{\Psi_0|\Psi_0}} = \frac{1}{Z} \sum_{i,j} \left[ e^{-\beta \left(\bar \HH (C_i) + \bar \HH (C_j)  \right) /2 } Q(C_i) \braket{C_j|C_i} \right]=  \braket{Q}_{\text{cl}} \, .  
\end{equation}
Thus the choice of normalization in
Eq.~(\ref{eq:Psi0-Norm}) is meaningful.

\bigskip

Let now $\HH$ be a Hamiltonian operator in the Hilbert space
$\mathscr{H}$ that admits $\ket{\Psi_0} $ as its unique zero energy
ground state,
\begin{equation}
  \HH \ket{\Psi_0} = 0 \, .
\end{equation}
We call $\HH $ the quantum Hamiltonian corresponding to $\bar
\HH$. Note in particular that $\beta $ is now just a parameter in
$\HH$. The \emph{quantum partition function} which describes the
quantum behaviour of the system contains on the other hand a new
coupling constant $\hbar$ and is defined as
\begin{equation}
  \mathscr{Z} = \Tr_{\mathscr{H}} ( e^{-\nicefrac{\HH}{\hbar}} ) \, .
\end{equation}
For a given operator $\hat Q$ the expectation value is defined as the
following path integral:
\begin{equation}
  \braket{ \hat Q } = \frac{1}{\mathscr{Z}} \Tr_{\mathscr{H}} ( \hat Q \, e^{-\nicefrac{\HH}{\hbar}} ) \, .
\end{equation}
The classical limit is obtained for $\hbar \to 0$ and using a saddle
point approximation one verifies that the main contribution to the
expectation value comes from the ground state, hence reproducing the
classical result:
\begin{equation}
  \braket{ \hat Q } =  \Tr_{\mathscr{H}} ( \hat Q \,  e^{-\nicefrac{\HH}{\hbar}} )  \xrightarrow[\hbar \to 0]{} \frac{\braket{\Psi_0 | \hat Q | \Psi_0}}{\braket{\Psi_0|\Psi_0}} = \braket{Q}_{\text{cl}}  \, .
\end{equation}

\bigskip

A geometric interpretation for the operator $\HH$ can be found in
terms of the classical configuration space. Consider the graph
obtained by associating each of the $C_i$ to a node and linking two
nodes with an edge weighted by $l_{ij} = \exp \left[ - \beta \left(
    \bar \HH (C_i) - \bar \HH(C_j) \right)\right]$. Define the
operator $\triangle_q$ on this graph as
\begin{equation}
  \left. \triangle_q \right|_{ij} =
  \begin{cases}
    \sum_{\braket{ k}_i} q^{(\bar \HH (C_i) - \bar \HH(C_k))/2 } & \text{if $i=j$} \\
    -1 & \text{if $i $ and $j$ are connected}\\
    0 & \text{otherwise.}
  \end{cases} 
\end{equation}
where $q = e^{-\beta}$. The notation $\braket{k}_i$ denotes
the sum running over all configurations $C_k$ adjacent to $C_i$. One
can easily verify that
\begin{equation}
  \triangle_q \ket{\Psi_0} = 0 \, .  
\end{equation}
Since this is the property we were looking for in the quantum
Hamiltonian, we take $\triangle_q=\HH$. There is some
arbitrariness in the choice of $\triangle_q$; in fact one could have
added longer range interactions and still preserved the
annihilation of $\ket{\Psi_0}$.

The $\triangle_q$ operator can be interpreted either as a $q$--deformed
Laplacian on the state graph or as the standard Laplacian plus a potential term:
\begin{equation}
  \label{eq:quantum-Hamiltonian-Laplacian}
  \HH\,f(u) = -J \sum_{\braket{v}_u}\left[ f(v) - f(u) \right] + J \sum_{\braket{v}_u}\left[ q^{(\bar \HH(u)-\bar{\HH}(v))/2} - 1 \right] f(u) = -J \left( \triangle + V_q(u) \right) f(u),
\end{equation}
where $J$ is the quantum energy scale (not to be confused with the
classical coupling $\beta$ of the classical Hamiltonian $\bar \HH$),
and $f(u)$ is a function defined on the graph. The choice of the minus
sign in front of the Hamiltonian is needed to ensure that
$\ket{\Psi_0}$ is the state of minimal energy. Note that in the $\beta
\to 0$ limit the potential term $V_q$ is proportional to the second
variation of the classical Hamiltonian.

\bigbreak

By following the above quantization description, the dynamics and a
time direction are added to the initial classical problem, since the
states in the Hilbert space satisfy the Schr\"odinger equation
\begin{equation}
  \imath\, \hbar\, \Deriv{}{t} \ket{\Psi (t)} = \HH \ket{\Psi(t)} \, .  
\end{equation}
This dynamics corresponds to a random walk in the space of configurations where we
introduce a (discrete) metric depending on the classical energy.

When this article was close to completion, we became aware
of~\cite{Castelnovo:2005} which has some overlap with the treatment given in this section.

\subsection{The quantum dimer model and quantum crystal melting}
\label{sec:qcrsytalmelting}

The quantum version of the dimer model was introduced in
1988~\cite{Rokhsar1988}. Given a dimer model on a bipartite graph, the
perfect matchings form a basis for the Hilbert space of the quantum
dimer model (\textsc{qdm}).  Its (quantum) Hamiltonian takes the form
\begin{equation}
  \label{eq:quantumham}
  \HH = -J \sum \left( \ketbra{\square}{\blacksquare} + \ketbra{\blacksquare}{\square} \right) + V \sum \left( \ketbra{\square}{\square} + \ketbra{\blacksquare}{\blacksquare} \right) \, ,
\end{equation}
where $\ket{\square}$ and $\ket{\blacksquare}$ represent the two
possible dimer configurations in a plaquette.

The first term is kinetic and runs over all pairs of perfect matchings
$\ketbra{\alpha}{\beta}$ which only differ by one elementary dimer
\emph{flip move} (plaquette move), in which a matching around a given
plaquette is turned by one position, thus exchanging edges covered by
a dimer with uncovered edges. The second, potential, term is diagonal
in the Hilbert space and counts the number of flippable plaquettes in
configuration $\alpha$.  The coupling constants $J$ and $V$ are
phenomenological parameters.  At the Rokhsar--Kivelson (\textsc{rk})
point $V=J$, the exact ground state wave function is an
equal--weighted superposition of all perfect matchings.

This construction can be generalized, as shown
in~\cite{Henley2003}. The dynamics of any discrete stochastic
classical model is described by the \emph{master equation}
\begin{equation}
  \label{eq:master}
  \frac{\mathrm{d}}{\mathrm{d}\tau}\, P_\alpha(\tau) = \sum_{\substack{\beta\\\beta\neq\alpha}}\left(W_{\alpha\beta} P_\beta(\tau) - W_{\beta \alpha} P_\alpha(\tau)\right),
\end{equation}
where $P_\alpha(\tau)$ is the probability to be in configuration
$\alpha$ at time $\tau$, and $W_{\alpha \beta}$ is the
\emph{transition rate} to state $\alpha$ if the system is in state
$\beta$.  It was shown in~\cite{Henley2003} (see
also~\cite{Castelnovo:2005}) that every discrete system fulfilling the
\emph{detailed balance} condition
\begin{equation}
  \label{eq:detailedBalance}
  W_{\beta\alpha}P^{(0)}_\alpha=W_{\alpha \beta}P^{(0)}_\beta
\end{equation}
has a \textsc{rk} point with the same mapping of the eigenfunctions to the classical dynamics.

The system of a melting crystal corner can be mapped to the problem
of stacking cubes in an empty corner of 3D space. The growth rules are the following. A cube can be added if three of its sides will touch either the wall or the faces of other cubes. This leads to a minimum energy configuration without any free--standing cubes. The partition
function of the melting crystal corner takes the following form:
\begin{equation}
  \label{eq:melting}
  Z = \sum_{3d\ \text{partitions }\alpha }q^{N(\alpha)}=\prod_{n=1}^\infty\,(1-q^n)^{-n},
\end{equation}
where $N(\alpha)$ is the number of boxes in the configuration $\alpha$ and the rightmost expression is the so--called \emph{MacMahon
  function}~\cite{MacMahon:1915}. The convergence of the above expression is only guaranteed for $0<q<1$, which we will assume henceforth.

\begin{figure}
  \begin{center}
    \begin{minipage}{.9\linewidth}
      \subfigure[The empty room perfect matching]{\includegraphics[width=.25\textwidth]{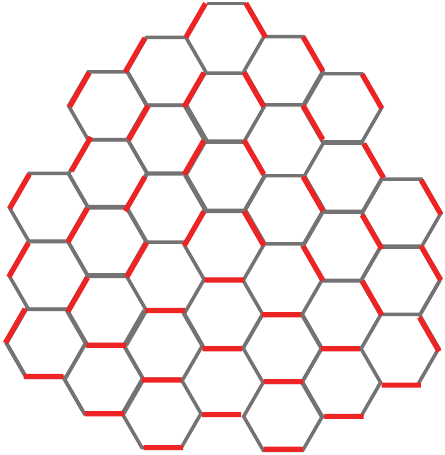}}
      \hfill
      \subfigure[After one plaquette move]{\includegraphics[width=.25\textwidth]{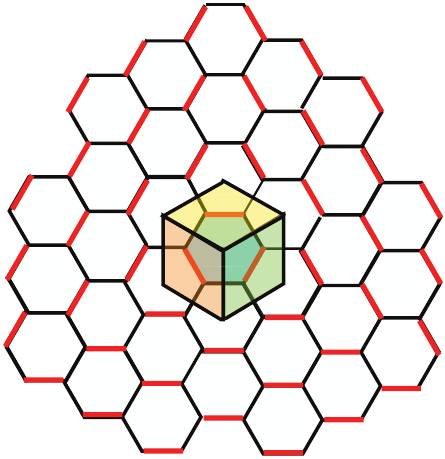}}
      \hfill
      \subfigure[After five plaquette moves]{\includegraphics[width=.25\textwidth]{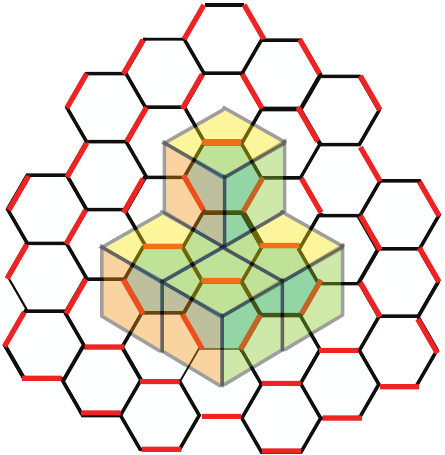}}
    \end{minipage}
  \end{center}
  \caption{Empty room perfect matching and matchings after flip moves}
  \label{fig:honey_cubes}
\end{figure}

The configurations of the melting crystal corner are in one--to--one
correspondence with the perfect matchings on an infinite hexagonal
lattice. There is a unique (up to translation) empty room perfect
matching which serves as the starting point. A plaquette move
corresponds to adding or removing a box~\cite{Okounkov:2003sp}, see
Figure~\ref{fig:honey_cubes}.  Nonetheless, the \textsc{qdm} explained
above does not correspond directly to the quantum system for the
melting crystal corner, since the number of boxes does not enter
anywhere. In the following, we derive the quantum Hamiltonian for the
crystal melting. Even though the following paragraph phrases
everything in the language of the three--dimensional crystal melting
problem, the procedure is completely general and applicable to any
dimension.

\begin{figure}
  \begin{center}
    \includegraphics[width=.9\textwidth]{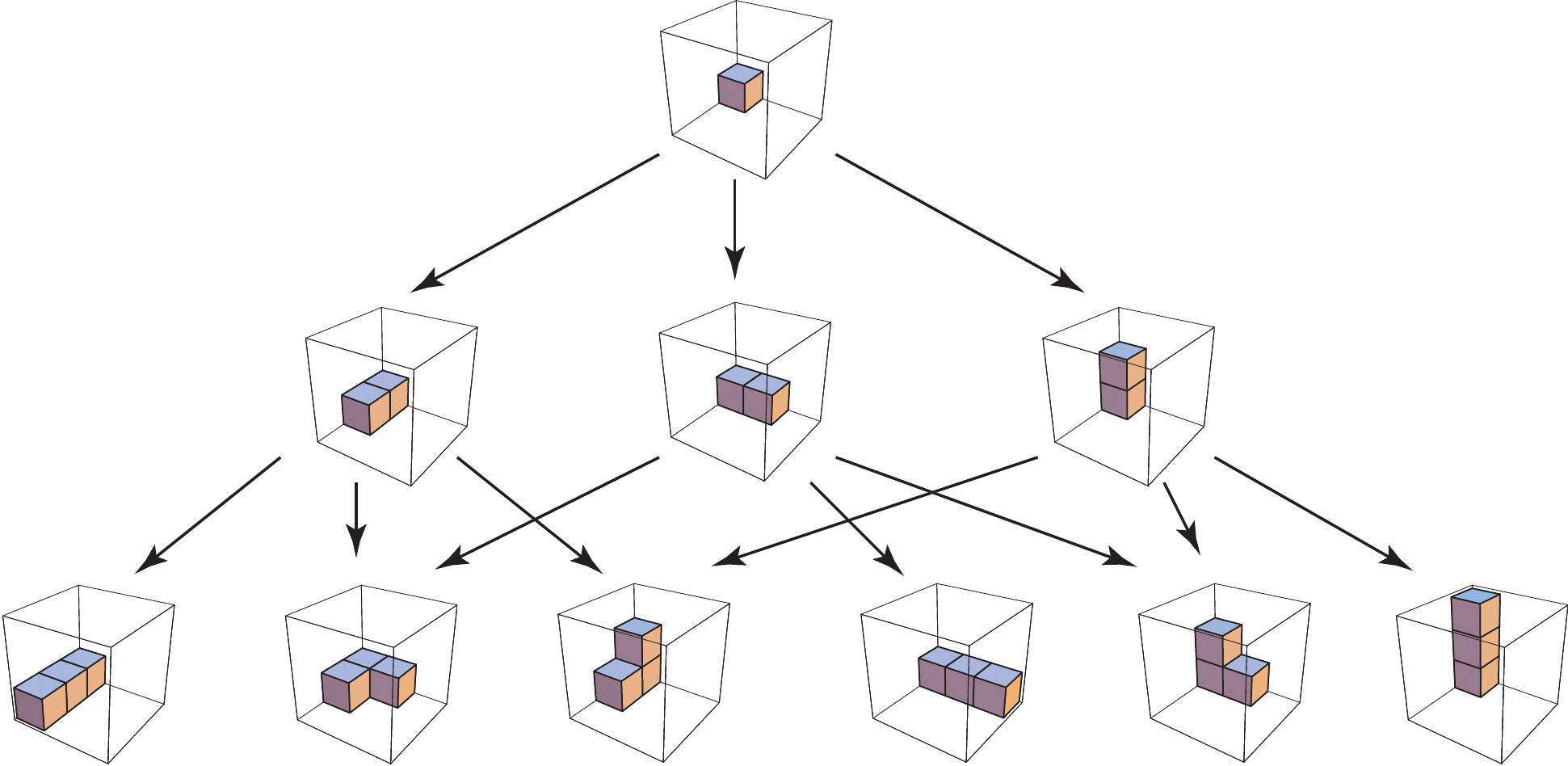}
  \end{center}
  \caption{State graph of the melting crystal corner for the first three levels}
  \label{fig:state-diagram}
\end{figure}

To describe the (classical) dynamics of the crystal melting process,
we express it as a random walk on the state graph (where each node
corresponds to a configuration and two nodes are joined by a directed
line if the endpoint can be reached from the initial one by adding a
box -- see Fig.~\ref{fig:state-diagram}). The dynamics is encoded in
a master equation of the type~(\ref{eq:master}).

We choose the following transition rates:
\begin{equation}
  W_{\alpha \beta} =
  \begin{cases}
    1/ \!\sqrt{q} & \text{if $N(\beta) = N(\alpha ) + 1$,} \\
    \sqrt{q} & \text{if $N( \beta) = N(\alpha ) - 1$,} \\
    0 & \text{otherwise.}
  \end{cases} 
\end{equation}
In other words,
\begin{equation}
  W_{\alpha \beta} = C_{\alpha \beta}\, q^{\left( N( \alpha ) - N (\beta)  \right)/ 2}  \, ,
\end{equation}
where $C_{\alpha \beta}$ is the adjacency matrix for the state
graph. One can verify that the (unique) stationary distribution for
$P_\alpha $ is
\begin{equation}
  P_\alpha^{(0)} = \frac{1}{Z}\, q^{N(\alpha)} \, ,  
\end{equation}
where, as in Eq.~\eqref{eq:melting},
\begin{equation}
  Z = \sum_\alpha q^{N(\alpha)} \, .  
\end{equation}
This is equivalent to saying that the system is in a Boltzmann
distribution with classical energy
\begin{equation}
  \bar \HH (\alpha ) = N( \alpha ) \, .  
\end{equation}
In fact, one can easily verify not only that $P_\alpha^{(0)}$ is an
equilibrium state, but that it also satisfies the detailed balance condition~(\ref{eq:detailedBalance}).
It is customary to define the exit rate from state $\alpha $ as
\begin{equation}
  W_{\alpha \alpha} = - \gamma_\alpha = - \sum_{\beta \neq \alpha } W_{\beta \alpha} = - \sum_{\beta \neq \alpha }   C_{\alpha \beta} q^{\left( N( \beta ) - N (\alpha )  \right)/ 2} = - \frac{1}{\sqrt{q}}\, \deg^+ (\alpha) - \sqrt{q}\, \deg^- (\alpha ) \, ,
\end{equation}
where $deg^\mp (\alpha)$ is the number of configurations connected to
$\alpha $ containing one block more (less) than $\alpha$. Note that in
this way, the evolution can be described by the vector equation
\begin{equation}
  \frac{\di }{\di t } \mathbf{P} (t) = \mathbf{W} \mathbf{P} (t) \, ,
\end{equation}
and the stationarity condition becomes
\begin{equation}
  \mathbf{W} \mathbf{P}^{(0)} = 0 \, .
\end{equation}
Instead of using the matrix $\mathbf{W}$, one can define the
symmetrized version
\begin{equation}
  \widetilde{W}_{\alpha \beta} = \left( P_\alpha^{(0)} \right)^{-1/2} W_{\alpha \beta}  \left( P_\beta^{(0)} \right)^{-1/2} 
\end{equation}
(no summation implied).  Explicitly, one finds
\begin{equation}
  \begin{cases}
    \widetilde{W}_{\alpha \beta} = C_{\alpha \beta} & \text{if $\alpha \neq \beta$} \\
    \widetilde{W}_{\alpha \alpha } = W_{\alpha \alpha } \, .
  \end{cases}
\end{equation}
Given the characteristic equation $\widetilde{W}_{\alpha \beta} \tilde
\phi_\beta^{(\lambda)} = \lambda \tilde \phi_{\alpha }^{(\lambda)} $
one can verify that the ground state is now represented by the vector
$\tilde \phi_{\alpha}^{(0)} = \sqrt{P_{\alpha}^{(0)}}$.

\bigskip 

Let us now switch to the $\ket{ \square}$--notation used in the
\textsc{qdm}. The ket $\ket{ \blacksquare}$ here represents a cube in
the configuration that can be removed, while $\ket{ \square}$
represents a place where a cube can be added.  Now,
$\widetilde{W}_{\alpha \alpha }$ can be written as
\begin{equation}
  \widetilde{W}_{\alpha \alpha } =  \braket{ \alpha | \left[ - \sum \sqrt{ q} \ketbra{\square}{\square} +  \frac{1}{\sqrt{q}} \ketbra{ \blacksquare}{\blacksquare} \right] | \alpha }  ,
\end{equation}
where the sum runs over all places where a cube can be added or removed in the configuration. Similarly, the non--diagonal terms can be written as
\begin{equation}
  \widetilde{W}_{\alpha \beta} = \braket{ \alpha | \left[ \sum \ketbra{\blacksquare}{\square} + \ketbra{\square}{\blacksquare} \right]| \beta} \, .  
\end{equation}
It is therefore natural to define the quantum Hamiltonian by
\begin{equation}
  \label{eq:QHamil}
  \HH =  - J \sum\ketbra{\blacksquare}{\square} +  \ketbra{\square}{\blacksquare} + V \sum  \sqrt{q}  \ketbra{ \square}{\square} + \frac{1}{\sqrt{q}} \ketbra{ \blacksquare}{\blacksquare} \, ,
\end{equation}
acting on the Hilbert space generated by the orthonormal basis of the configurations $\ket{\alpha}$. Notice that at the \textsc{rk} point $V/J = 1$,
this coincides with $\widetilde{\mathbf{W}}$:
\begin{equation}
  \HH = - J \,\widetilde{\mathbf{W}} \, .
\end{equation}
Being proportional, $\HH $ and $\widetilde{\mathbf{W}}$ share the same
eigenvectors with proportional eigenvalues:
\begin{equation}
  \HH_{\alpha \beta}\, \tilde \phi^{(\lambda)}_\beta = -J  \lambda \, \tilde \phi^{(\lambda)}_\alpha \, .
\end{equation}
In particular, the ground state $E_0 = 0$ corresponds to
\begin{equation}
  \tilde \phi_\alpha^{(0)} = \sqrt{ P_\alpha^{(0)}} = \frac{1}{\sqrt{Z}}\, q^{N(\alpha)/2}\, ,  
\end{equation}
which has to be interpreted now in the sense of a quantum superposition:
\begin{equation}\label{eq:ground1}
  \ket{\text{ground}} = \sum_\alpha  q^{N(\alpha)/2} \ket{\alpha } \, ,
\end{equation}
where $\ket{\alpha }$ is the state corresponding to the perfect
matching $\alpha $. Note that we changed the normalization for the
wave function as in Eq.~(\ref{eq:Psi0-Norm}), so that
\begin{equation}
  \braket{\text{ground} | \text{ground}} = \sum_\alpha q^{N(\alpha)} = Z \, .  
\end{equation}

As as consequence of the detailed balance condition in
Eq.~(\ref{eq:detailedBalance}), the ground state is frustration free,
\ie each local interaction is minimized separately.


%% file: OneDimension.tex
\section{The one dimensional problem}
\label{sec:one-dimens-young}

As a warm--up we briefly consider the one--dimensional analog to the
crystal melting problem, \ie a system where a set of blocks is added
or removed along a line with integer lattice points starting from
$n=0$. The states are labelled by an integer number $\ket{n}$ and the
quantum Hamiltonian acts on them as follows (note that we placed
ourselves at the $V=J$ point):
\begin{equation}
  \begin{cases}
    \HH \ket{0} = -J \left[ \ket{1} - q^{1/2} \ket{0} \right] \, , \\
    \HH \ket{n} = -J \left[ \ket{n+1} + \ket{n-1} - \left( q^{1/2} + q^{-1/2} \right) \ket{n} \right]\, , & \text{$\forall\ n > 0$ \, .}
  \end{cases}
\end{equation}
As shown in the previous section, the ground state is
\begin{align}
  \ket{\Psi_0} = \sum_{n=0}^\infty q^{n/2} \ket{n} \, , && Z=\braket{\Psi_0 | \Psi_0 } = \sum_{n=0}^\infty q^n = \frac{1}{1-q} \, .
\end{align}
For $n>0 $, the Hamiltonian can be written in terms of the Laplacian
on the line as
\begin{equation}
  \HH =  -J \triangle + m^2 \, ,
\end{equation}
where
\begin{align}\label{eq:gap1d}
  \triangle \ket{n} &= \ket{n+1} + \ket{n-1} - 2 \ket{n} \, ,\\
  m^2 &= J \left( q^{1/2} + q^{-1/2} - 2 \right) \, .
\end{align}
Hence we can immediately read off the value for the mass gap.  In fact
the system is easily solvable and one finds the eigenvectors
\begin{equation}
  \ket{\Psi(k)} = \sum_{n=0}^\infty \left( \sin (k \, n) + q^{1/2} \sin (k \left( n + 1 \right))  \right) \ket{n},   \quad  k \neq 0\,, 
\end{equation}
with energies
\begin{equation}
  \HH \ket{\Psi(k) } =  J \left( q^{1/2} + q^{-1/2} - 2 \cos k \right) \ket{\Psi(k)} \,  .
\end{equation}
This set of eigenstates does not contain the first excited state 
$\ket{\Psi_1}$ which would correspond to $k=0$. It
satisfies
\begin{equation}
  \HH \ket{\Psi_1} = m^2 \ket{\Psi_1} \, .  
\end{equation}
This means that $\ket{\Psi_1}$ is harmonic:
\begin{equation}
  \triangle \ket{\Psi_1} = 0 \, .  
\end{equation}
In one dimension, this is easily solved and using the equation for
$\ket{0}$ as initial condition one finds
\begin{equation}
  \ket{\Psi_1 } = \sum_{n=0}^\infty \left( -n +  q^{1/2}\left( n + 1\right) \right) \ket{n} \, .  
\end{equation}


%% file: FermionFormalism2d.tex
\section{The two--dimensional problem: quantum partitions}
\label{sec:2d_partitions}

After having discussed the simplest version of the problem in the last
section, we now turn to the study of the growth of a two--dimensional
crystal. When using the picture of squares being added to an empty
corner of the plane, the growth rules are the following: a square can
only be added when two of its sides will touch either the wall or the
side of another square. It is easy to see that in this case the
allowed crystal configurations are in one--to--one correspondence with
random partitions.  A \emph{partition} of $n$ is a non--increasing
finite sequence of positive integers whose sum is $n$.  We can
visually represent a partition by its (Ferrers) diagram, for which we
adhere to the Russian tradition, see Figure~\ref{fig:2d}, where the
corner of the plane is rotated by 135 degrees. The theory of
partitions is at the same time an old subject of mathematics with many
important results dating back to Euler, and one which has seen major
progress in the last fifty years (see \emph{e.g.}~\cite{Andrews} for
an overview).

Let $\set{C_i}$ be the set of all integer partitions (or equivalently
Young diagrams). Since we are interested in the size of a partition,
it is natural to define the ``classical'' Hamiltonian
\begin{equation}
  \bar \HH [ C_i] = \text{\# of boxes of the partition $C_i$.}  
\end{equation}
The classical partition function for 2d partitions is given by
\begin{equation}\label{eq:partfn2d}
 Z^{\text{2d}} = \sum_{C_i}\,q^{\bar \HH [ C_i]}=\prod_{n=1}^\infty\,(1-q^n)^{-1}.
\end{equation}

Using the procedure detailed in Sec.~\ref{sec:class-vs.-quant} we
define the graph of configurations whose nodes are in one--to--one
correspondence with the $C_i$ and where a line is drawn between two
partitions if they differ by only one square. The Hamiltonian retains
the form given in Eq.~(\ref{eq:QHamil}),
\begin{equation}
  \label{eq:QHamil2}  
  \HH =  - J \sum\ketbra{\blacksquare}{\square} +  \ketbra{\square}{\blacksquare} + V \sum  \sqrt{q}  \ketbra{ \square}{\square} + \frac{1}{\sqrt{q}} \ketbra{ \blacksquare}{\blacksquare} \, ,
\end{equation}
acting on the Hilbert space generated by the orthonormal basis of the
configurations $\ket{C_i}$. The ket $\ket{ \blacksquare}$ now
represents a square that can be removed from the partition, while
$\ket{ \square}$ represents a place where a square can be added. The
sum runs over all places where a square can be added or removed in the
configuration.

The 2d case of random partitions is especially interesting for us
because of the existence of a map between 2d partitions and the
Neveu--Schwarz sector of the complex fermionic oscillator. It enables
us to express the quantum Hamiltonian~(\ref{eq:QHamil2}) in terms of
fermionic operators.

\subsection{Fermion operator formalism for 2d partitions}
\label{sec:fermion2d}

\begin{figure}
  \begin{center}
    \begin{minipage}{.9\linewidth}
      \subfigure[Empty partition]{\includegraphics[width=.45\textwidth]{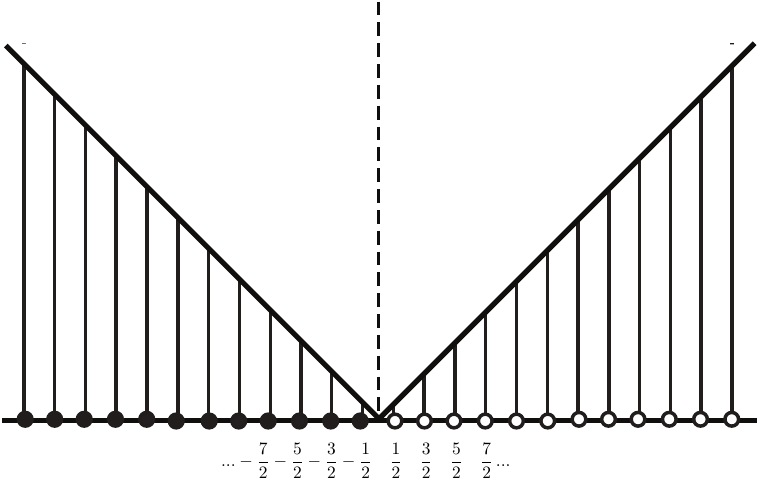}}
      \hfill
      \subfigure[Partition with one box]{\includegraphics[width=.45\textwidth]{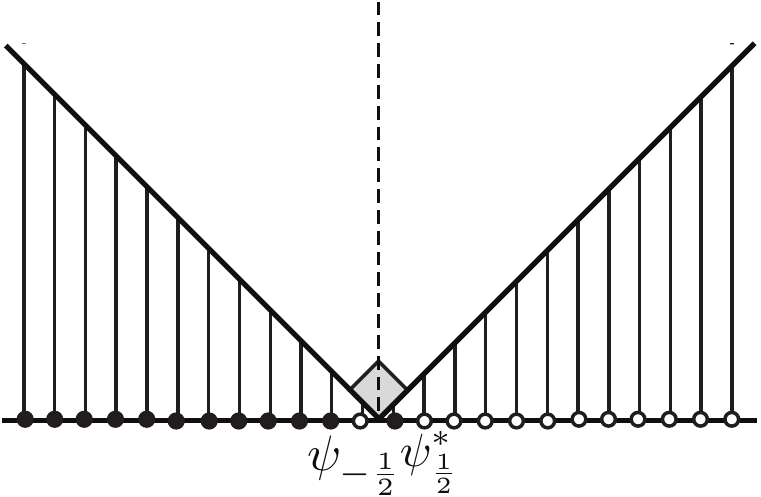}}
    \end{minipage}
  \end{center}
  \caption{Empty partition and partition with one square with projection to 1d fermions}
  \label{fig:empty_one}
\end{figure}

\begin{figure}
  \begin{center}
    \includegraphics[width=.5\textwidth]{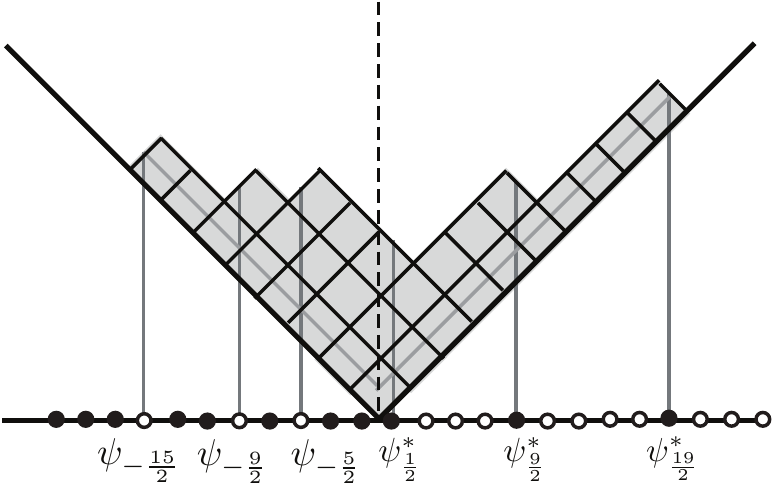}
  \end{center}
  \caption{Two--dimensional partition corresponding to the state
    $\psi_{\frac{19}{2}}^* \psi_{-\frac{15}{2}}\psi_{\frac{9}{2}}^*
    \psi_{-\frac{9}{2}}\psi_{\frac{1}{2}}^* \psi_{-\frac{5}{2}}
    \ket{\text{half}} $.}
  \label{fig:2d}
\end{figure}

In this section, we will explain the map from partitions to fermions
mentioned above. The empty partition is depicted in
Figure~\ref{fig:empty_one}.(a). The horizontal line shows the
projection to the 1d fermion problem, where the fermions sit at
half--integer positions. All the negative positions are filled,
therefore the empty partition corresponds to the filled Fermi sea. The
fermion operators obey the usual anti--commutation relations
\begin{equation}
\{\psi_n,\psi_m^*\}=\delta_{n,m},\quad n,m\in\mathbb{Z}+1/2,
\end{equation}
all other anti--commutators zero, and the annihilation operators
annihilate the vacuum,
\begin{equation}
\psi_n\ket{0}=0\quad \forall n\in\mathbb{Z}+1/2.
\end{equation}
Note that the vacuum $\ket{0}$ corresponds to the bottom of the Fermi
sea.
In terms of fermions, the empty configuration (no squares) is
represented by
\begin{equation}
  \label{eq:half}
  \psi^*_{-\infty}\dots\psi^*_{-\frac{3}{2}}\psi^*_{-\frac{1}{2}}\ket{0}=\ket{\text{half}},
\end{equation}
where each NW--SE diagonal line corresponds to a creator $\psi^*_{a}$.
The state $\ket{\text{half}}$, corresponding to the filled Fermi sea
(a picture also familiar from matrix models), obeys
 \begin{equation}
  \label{eq:FermiSea}
  \psi^*_{-n}\ket{\text{half}}=0,\quad \psi_{n}\ket{\text{half}}=0,\quad n>0.
\end{equation}
The partition with one square, see Figure~\ref{fig:empty_one}.(b), 
corresponds to
\begin{equation}
  \label{eq:one}
  \psi_{-\frac{1}{2}}\psi^*_{\frac{1}{2}}\ket{\text{half}}=\psi^*_{-\infty}\dots\psi^*_{-\frac{3}{2}}\psi^*_{\frac{1}{2}}\ket{0},
\end{equation}
which corresponds to an annihilator for the SW--NE line at
$-\tfrac{1}{2}$ and a creator for the NW--SE line at $\tfrac{1}{2}$.
On the projection to the horizontal line, the fermion at position
$-\tfrac{1}{2}$ was annihilated and a fermion was created at
$\tfrac{1}{2}$, \emph{i.e.} the fermion at $-\tfrac{1}{2}$ hopped over
to $\tfrac{1}{2}$. Adding squares to the partition corresponds to
fermions hopping to the right, in this sense this can be seen as a
diffusion process in which every position can be occupied only by one
particle, (\emph{i.e.} it is \emph{exclusive}).
Figure~\ref{fig:2d} shows a partition with the corresponding occupied and
free positions shown in the projection to the horizontal line.  The
Hamiltonian~(\ref{eq:QHamil2}) becomes now
\begin{equation}
  \label{eq:Hamil2d}
  \HH_{\text{2d}} = -J \left(\sum_{m \in \setZ + 1/2}\psi^*_{m+1}\psi_m+\psi^*_m\psi_{m+1} - \frac{V}{J} \sqrt q\,n_m \left( 1 - n_{m+1} \right) - \frac{V}{J} \frac{1}{\sqrt q}\,n_{m+1} \left( 1 - n_m \right)\right) \, ,
\end{equation}
where $n_m=\psi_m^*\psi_m$ is the fermion number operator. The first
two terms are the kinetic hopping terms, while the last two terms are
the potential terms, counting the number of possibilities to hop right
and left per configuration. The different weights for hopping right
and left ($\sqrt q$ and $1/\sqrt q$) make the system into an
\emph{asymmetric} diffusion process.  The fermion number $n_f$ is
conserved in this system, therefore states with differing fermion
numbers belong to different superselection sectors ${\cal S}_{n_f}$ of
the Hamiltonian.
%
%

Let us now go to the \textsc{rk} point $V = J$. The unique ground state,
satisfying $\HH \ket{\text{ground}} = 0 $, is written as in
Eq.~(\ref{eq:ground1}):
\begin{equation}
  \label{eq:ground_2d}
  \ket{\text{ground}}=\lim_{d\to\infty}\left[\sum_{\substack{a_1 < a_2 < \dots < a_d \\  b_1 < b_2 < \dots < b_d}} q^{\frac{1}{2}\sum_{i=1}^d(a_i+b_i)} \prod_{i=1}^d \psi_{a_i}^* \psi_{-b_i} \right] \ket{\text{half}} \, ,
\end{equation}
where we notice that the number of squares in a given
partition $\mu$ is given by
\begin{equation}\label{eq:nusq}
|\mu|=\sum_{i=1}^d(a_i+b_i).
\end{equation}


\subsection{The Heisenberg \textsc{xxz} spin chain}
\label{sec:XXZ}

The Hamiltonian in Eq.~(\ref{eq:Hamil2d}) can be recast into a
different form, which is more familiar in statistical physics. If we
express the creation and annihilation operators in terms of the usual
Pauli matrices $\sigma_m^k, \, k=1,2,3$ at position $m$,
Eq.~(\ref{eq:Hamil2d}) becomes\footnote{Note that our notation differs
  form the one generally used in condensed matter literature, where
  not $\sqrt q$, but $q'$ is used which is related to $q$ by $q'=\sqrt
  q$.}
\begin{equation}
  \label{eq:Heisenberg}
  \begin{split}
    \HH_{\text{2d}} =& -\frac{J}{2} \sum_{m\in \setZ + 1/2} \left[\,\sigma_m^1\sigma_{m+1}^1+\sigma_m^2\sigma_{m+1}^2 + \frac{V}{2J} \left( \sqrt q  +\frac{1}{\sqrt q}\right)  \sigma_m^3\sigma_{m+1}^3 + \right.\\
    &\left.+\frac{V}{2J} \left( \sqrt q - \frac{1}{\sqrt q} \right)\, \left( \sigma_m^3 - \sigma_{m+1}^3 \right) - \frac{V}{2J} \left( \sqrt q  +\frac{1}{\sqrt q} \right) \right] \, .
  \end{split}
\end{equation}
Going to the \textsc{rk} point $V/J=1$, this is precisely the
Hamiltonian for the Heisenberg \textsc{xxz} spin
chain~\cite{Alcaraz3} with anisotropy parameter
\begin{equation}
  \Delta = \frac{\sqrt q  +\frac{1}{\sqrt q}}{2}\, .
\end{equation}
Note that the term proportional to $\left( \sigma_m^3 - \sigma_{m+1}^3
\right)$ is the only term in Eq.~(\ref{eq:Heisenberg}) not invariant
under the exchange $q \to 1/q$. It is a boundary contribution
vanishing for the case of the infinite chain. In fact one could have
noticed this symmetry before since for any given 2d partition one
square more can be added than removed. It follows that the asymmetry
between $q$ and $1/q$ is only proportional to a trivial constant
term. This implies also that we can restrict our attention in this
case to $V=J$ without loss of generality.

Since for the partitions, we are interested in the case $0<q<1$, we
have $1<\Delta<\infty$, which means that we are in the ferromagnetic
regime. The case $\Delta=1$ corresponds to the isotropic Heisenberg or
\textsc{xxx} model. For $\Delta=\infty$, this is the one--dimensional
Ising model.  The \textsc{xxz} model is one of the best studied
systems in statistical mechanics and can be solved using the Bethe
ansatz~\cite{Bethe:1931hc}.  It was shown to correspond to the
six--vertex or ice--type model.  That the finite \textsc{xxz} chain
admits the quantum group $U_{\sqrt{q}}[SU(2)]$ as a symmetry was first
pointed out by Pasquier and Saleur~\cite{Pasquier:1990}. A large part
of the literature is devoted to the anti--ferromagnetic case or the
case $-1<\Delta<1$, but a fair amount of results exists on the
ferromagnet, which is collected in~\cite{Starr}.

\begin{figure}
  \centering
  \includegraphics[width=.5\textwidth]{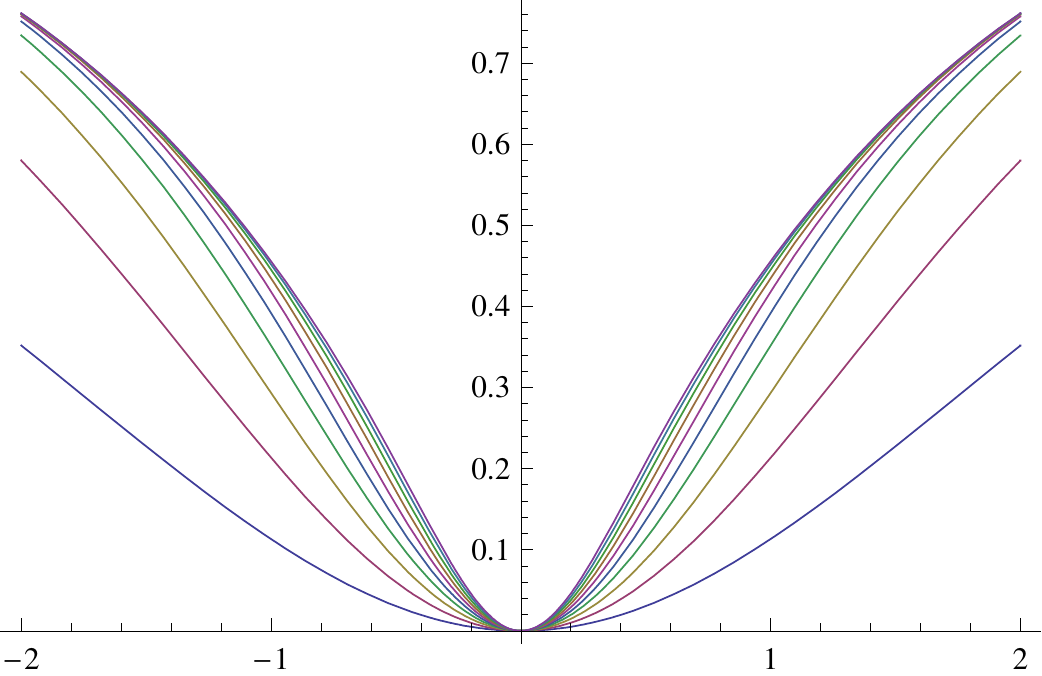}
  \caption{Energy levels for the droplet states as a function of $\beta$ for different values of the spin deviation as in Eq.~(\ref{eq:droplet}).}
  \label{fig:droplet-energies}
\end{figure}

We are specifically interested in the infinite volume chain. It admits
four types of zero energy ground states. Two are translation invariant
and correspond to all spins up and all spins down. The other two are
the kink and the anti--kink~\cite{Alcaraz,Gottstein}.  This list of
ground states was shown to be complete. These ground states are
furthermore frustration free, \ie\ they minimize each of the
next--neighbour interactions separately. Over each of these ground
states, a spectral gap exists with value~\cite{Koma}
\begin{equation}
  \label{eq:gap}
  \gamma = -J \left( q^{1/2} + q^{-1/2} - 2 \right) \, .
\end{equation}
It was furthermore shown in~\cite{Nachtergaele} that droplet
excitations exist over all four ground states. In our normalization,
their energies are~
\begin{equation}
  \label{eq:droplet}
  E(n) = -J \left( \frac{1}{q^{1/2}} - q^{1/2} \right) \frac{1-q^{n/2}}{1+q^{n/2}} \, ,
\end{equation}
where $n$ is the number of spins that deviate from the ground state
and $g = - \log q$. For $n=0$, $E(0)=0$, while for $n=1$, the mass gap
Eq.~(\ref{eq:gap}) is reproduced. These energy levels are symmetric
under the exchange $q \to 1/q$ as shown in
Fig.~\ref{fig:droplet-energies}.

\subsection{Correlation functions for the kink states}
\label{sec:corr-funct-kink}

Since the partitions correspond to the case of the kink centered in 0,
we concentrate on the kink states.  The kinks form an infinite family
of ground states interpolating between spin up at $-\infty$ and spin
down at $+\infty$. These kinks are the lowest energy eigenstates for
the superselection sectors with fixed particle number into which the
Hilbert space separates (note that $\comm{H, \sum_m \sigma^3_m} = 0$,
which means that the action of the Hamiltonian does not change the
particle number). 

\subsubsection{Grand canonical ground state}
\label{sec:grand-canon-ground}

A useful representation for the kink state can be
obtained in terms of the grand canonical ground
state
\begin{equation}
\label{eq:kink}
  \Psi (z) = \prod_{n=0}^\infty \left( 1 + z \sqrt{q}^{n+1/2} \psi^\ast_n \right) \ket{0} = \bigotimes_{n = 0 }^\infty  \binom{1}{z \, \sqrt{q}^{n + 1/2}} \, ,
\end{equation}
which represents a kink centered in $x_0 = - \log \abs{z} / \log q
$. Note that we shifted the spin chain such that the leftmost spin is at
$n = 1/2$. The kink is exponentially localized~\cite{Bach:2000}. In
fact if one considers the magnetization profile $m_{\Psi(z)}(x) =
\braket{\Psi(z) | \sigma^3_x | \Psi(z)}$, one finds that
\begin{equation}
  \begin{cases}
    \frac{1}{2} - m_{\Psi(z)}(x) \leq q^{\abs{x - x_0}} & \text{if $x < x_0$} \\
    \frac{1}{2} + m_{\Psi(z)}(x) \leq q^{\abs{x - x_0}} & \text{if $x > x_0$.}     
  \end{cases}
\end{equation}
For $q \to 0$, the system becomes an Ising model, where the kink is
localized in a single point, \ie\ the magnetization profile becomes a step
function centered in $x=x_0$. For $q = 1$, where the symmetry
is the usual $SU(2)$, the system is the \textsc{xxx} spin chain which
only admits translation--invariant ground states (one could say that
the kink is completely delocalized).

The grand--canonical kink~(\ref{eq:kink}) can be seen as a generating
function for the kinks in the $N$--particle sectors:
\begin{equation}
  \ket{\Psi (z)} =\lim_{L\to \infty} \sum_{N=0}^L \ket{\Psi_N} z^N \, .
\end{equation}
In particular, our ground state corresponds to the half--filled
infinite spin chain, \ie the kink centered in the middle of the chain
and is therefore precisely $\ket{\Psi_{L/2}}$.

Our strategy to compute correlation functions is the following: given
an operator $\mathscr{O}$ that does not depend on $z$ and leaves the
fermion number invariant, one can expand the correlator on the
grand--canonical kink as
\begin{equation}
  \braket{\Psi(z)| \mathscr{O} | \Psi(z)} = \sum_{N=0}^\infty z^{2N} \braket{ \Psi_N | \mathscr{O} | \Psi_N} \, ,
\end{equation}
where we used the fact that the $\ket{\Psi_N}$ are orthogonal since
they describe states with different particle numbers. It follows that
the correlation function on the $N$--particle kink is given by
\begin{equation}
  \braket{\mathcal{O}}_N = \frac{\braket{\Psi_N|\mathscr{O} | \Psi_N} }{\braket{\Psi_N | \Psi_N}} = \frac{1}{\braket{\Psi_N|\Psi_N}}  \oint \frac{\di z}{2 \pi \imath} \frac{\braket{\Psi(z) | \mathscr{O}| \Psi(z)}}{z^{2N +1}} \, .
\end{equation}
This is particularly useful when $\mathscr{O}$ does not mix
contributions from different points in which case the correlator on the
grand--canonical kink can be found easily.

As a first example consider $\mathscr{O} = 1$, \ie  the norm
\begin{equation}
  \braket{\Psi(z)| \Psi(z)} = \prod_{n=0}^\infty \left( 1 + z^2 q^{n+1/2} \right) = ( -w\, q^{1/2} ; q )_\infty \, , 
\end{equation}
where we used the notation of the $q$--shifted
factorials and $w = z^2$. Using the $q$--binomial theorem,
\begin{equation}
  ( - w\, q^{1/2}; q ) = \sum_{N=0}^\infty \frac{q^{N^2/2}}{(q;q)_N} w^N \, ,  
\end{equation}
we immediately find that the norm of an $N$--particle kink is
\begin{equation}
  \braket{\Psi_N| \Psi_N} = \frac{q^{N^2/2}}{(q;q)_N} = q^{N^2/2} \prod_{k=1}^N \frac{1}{\left(1 - q^k \right)} \, ,  
\end{equation}
where the factor $q^{N^2/2}$ takes the shift of the spin
chain into account.

\subsubsection{One--point function}
\label{sec:one-point-function}

\begin{figure}
  \begin{center}
    \begin{minipage}{.8\linewidth}
      \subfigure[Magnetization]{\includegraphics[width=.4\textwidth]{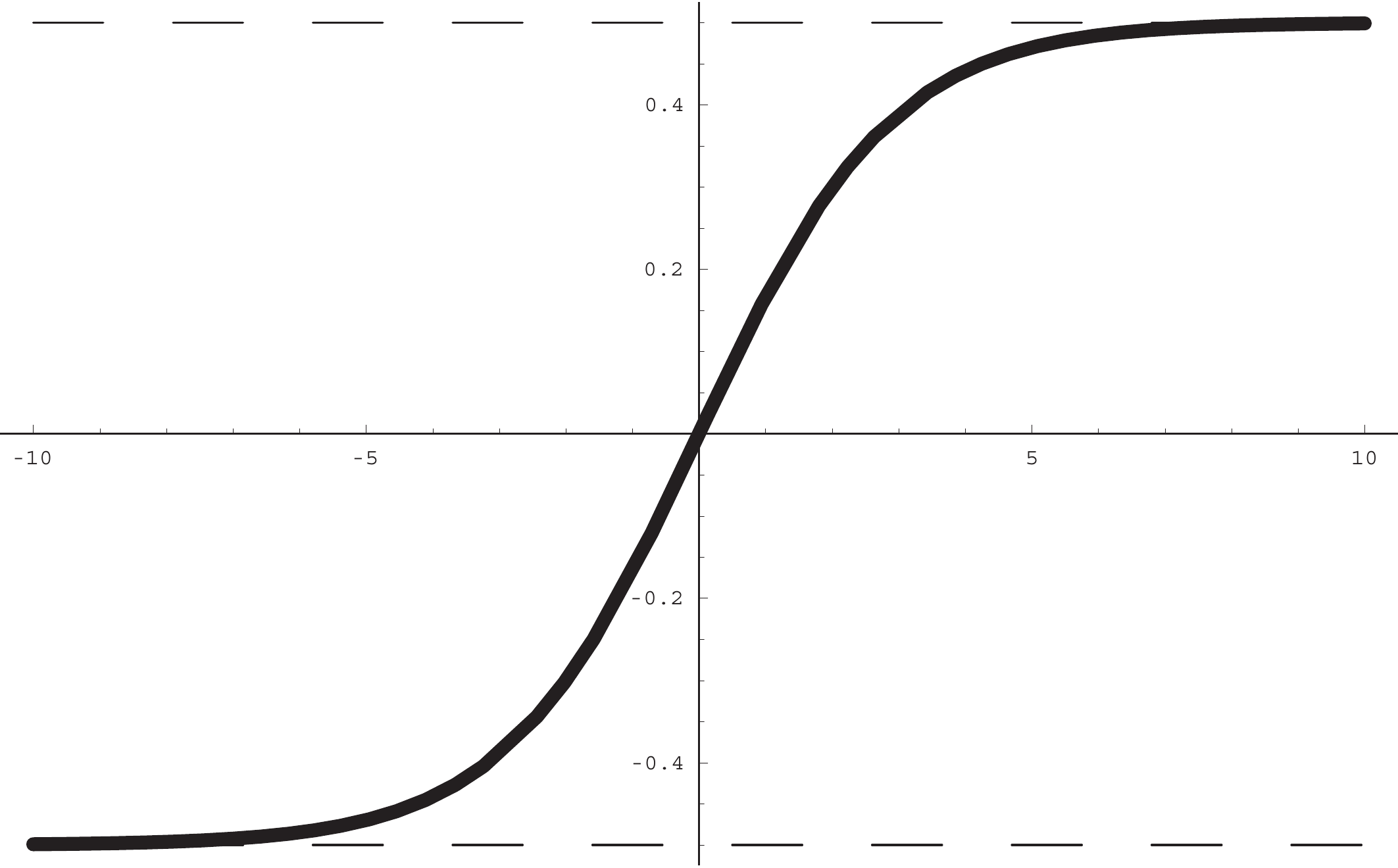}}
      \hfill
      \subfigure[Limit shape]{\includegraphics[width=.4\textwidth]{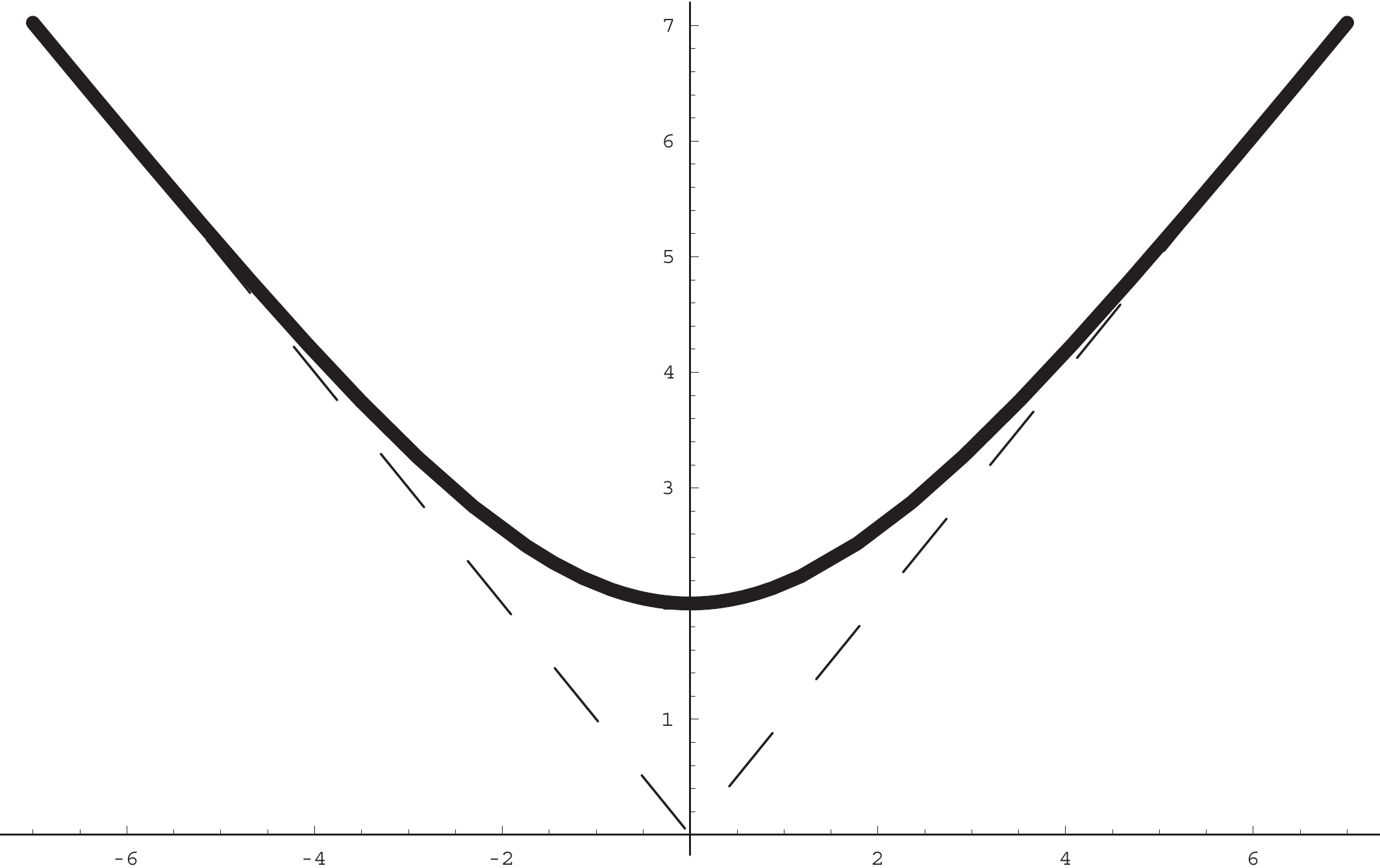}}
    \end{minipage}
  \end{center}
  \caption{Magnetization (a) and limit shape (b) for the ground state of
    the two--dimensional partition.}
  \label{fig:magnetization-limitshape}
\end{figure}

In general, other quantities of interest such as $n$--point
correlation functions can be evaluated for $\Psi(z)$ and then expanded
in a series of $w$ to get the contribution of the $N$--particle kink.
As an example, let $\mu(x)$ denote the profile of a partition $\mu$,
(\emph{i.e.} a piecewise $\mathcal{C}^\infty$ function). The map to
fermionic states is such that if $\mu'(\bar x) = -1$, there is a spin
up at position $\bar x$ in the chain, while if $\mu'(\bar x) = 1$,
there is a spin down at $\bar x$. Consider the operator $\sigma_{\bar
  x}^3$ acting on the point $\bar x$,
\begin{align}
  \sigma^3_{\bar x} \ket{ \uparrow} =  \frac{1}{2} \ket{\uparrow} \, , &&  \sigma^3_{\bar x} \ket{ \downarrow} =  -\frac{1}{2} \ket{\downarrow} \, .  
\end{align}
If we define the magnetization for a state $\ket{\Phi}$ by
\begin{equation}
  m_\Phi (x) = \frac{\braket{\Phi| \sigma^3_x | \Phi}}{\braket{\Phi|\Phi}} \, , 
\end{equation}
it is clear that
\begin{equation}
  m_\Phi (x) = - \frac{1}{2} \Deriv{}{x} \mu_\Phi (x) \, ,
\end{equation}
where $\mu_\Phi (x)$ is the shape of the partition associated to
$\ket{\Phi}$. The evaluation of the one--point function for the
generating function of the kinks yields
\begin{equation}
  \braket{\Psi(z) | \sigma_x^3 | \Psi(z)} = \frac{1 - w\, q^{x+1/2}}{2\left(1 + w \,q^{x+1/2}\right)} \prod_{n=0}^\infty \left( 1 + z^2 q^{n+1/2} \right) \, .
\end{equation}
By developing in a series (see App.~\ref{sec:calculations}) one can show that
the contribution of the $N$--particle kink is given by
\begin{equation}
   m_N (x) = \braket{\sigma^3_x}_N = \frac{\braket{\Psi_N | \sigma^3_x | \Psi_N}}{\braket{\Psi_N|\Psi_N}} =  -\frac{1}{2} +  \sum_{k=0}^N (q^{-N} ;\, q)_k \,q^{k \left( x + 1 \right)} \, .  
\end{equation}
Shifting the center and sending $N \to \infty$ one finds the \emph{exact}
expression for the shape of the kink:
\begin{equation}
  m_\infty(x) = -\frac{1}{2} +  \sum_{k=0}^\infty \left( - 1 \right)^k q^{\binom{k}{2}} q^{k x} \, .  
\end{equation}
To study the $q\to 1^-$ limit one can introduce a $q$--difference
equation satisfied by $m_\infty(x)$, see
App.~\ref{sec:calculations}. It follows that the magnetization profile
is
\begin{equation}
  m (x) =  -\frac{1}{2} \tanh (\frac{x}{2}) \, .  
\end{equation}
The shape of the partition is hence given by
\begin{equation}
  \mu_\phi (x) = - 2\int \di x \; m_\phi (x) = 2 \log ( 2 \cosh( \tfrac{x}{2} ))
\end{equation}
or, in a more symmetric form (take $\mu_\phi (x)=:y$),
\begin{equation}
  e^{(x-y)/2} + e^{-(x+y)/2} = 1 \, ,  
\end{equation}
see Fig.~\ref{fig:magnetization-limitshape}.  This reproduces the
limit shape for 2d partitions given in~\cite{Vershik:1996aa}.

\subsubsection{Two--point function}
\label{sec:two-point-function}

Given the factorized form of $\Psi(z)$ in Eq.~\eqref{eq:kink}, one can
easily evaluate two--point functions. In particular,
\begin{equation}
  \braket{\Psi(z) | \sigma_{x_1}^3 \sigma_{x_2}^3 | \Psi(z)} = \frac{\left(1 - w \, q^{x_1 + 1/2} \right)\left(1 - w\, q^{x_2 + 1/2} \right)}{4\left(1 + w\, q^{x_1 + 1/2} \right)\left(1 + w\, q^{x_2 + 1/2} \right)}  (-w\, q^{1/2} ; q)_\infty \, .
\end{equation}
Let $q^{x_i} = \zeta_i$. We have the expansion
\begin{equation}
  \frac{\left(1 - w\, \zeta_1 \right)\left(1 - w\, \zeta_2 \right)}{\left(1 + w\, \zeta_1 \right)\left(1 + w\, \zeta_2 \right)} = 1 + 2 \frac{\zeta_1 + \zeta_2}{\zeta_1 - \zeta_2}\sum_{k=0}^\infty (-1)^k w^k \left( \zeta_1^k - \zeta_2^k  \right) \, .
\end{equation}
We can now repeat the same construction as for the one--point function and find the contribution of the $N$--particle kink:
\begin{equation}
  \braket{\sigma^3_{x_1} \sigma^3_{x_2} }_N = \frac{\braket{\Psi_N|\sigma^3_{x_1} \sigma^3_{x_2}| \Psi_N}}{\braket{\Psi_N| \Psi_N}} = \frac{1}{4} +  \frac{1+q^{x_2-x_1}}{2 \left(1-q^{x_2-x_1}\right)} \sum_{k=0}^N (q^{-N}; q)_k q^{Nk} \left( q^{k x_1} - q^{k x_2} \right) \, .
\end{equation}
In the $N\to \infty$ limit, this becomes
\begin{equation}
  \braket{\sigma^3_{x_1} \sigma^3_{x_2}}_\infty = \frac{1}{4} + \frac{1+q^{x_2-x_1}}{2\left(1-q^{x_2-x_1}\right)} \sum_{k=0}^\infty \left( -1 \right)^k q^{\binom{k}{2}} \left( q^{kx_1} - q^{kx_2} \right) \, .
\end{equation}
To take the $q \to 1^-$ limit, we define $f(x)$ as in Eq.~\eqref{eq:limit-fq} and, after the rescaling $x_i \to - x_i / \log(q)$,
\begin{multline}
  \braket{\sigma^3_{x_1} \sigma^3_{x_2}}_\infty \sim \frac{1}{4} -
  \frac{1}{2} \coth(\frac{x_1-x_2}{2}) \left(
    \frac{e^{x_1}}{1+e^{x_1}} - \frac{e^{x_2}}{1+e^{x_2}}\right) = \\ =
  \frac{1}{4} \tanh ( \frac{x_1}{2} ) \tanh ( \frac{x_2}{2} ) =
  \braket{\sigma^3_{x_1} }_\infty \braket{ \sigma^3_{x_2}}_\infty  \, ,
\end{multline}
which shows that the two--point correlation function factorizes. The
same holds true for the $n$--point functions as we show in the
following section.

\subsubsection{The $n$--point function}
\label{sec:n-point-function}

The calculation of the $n$--point function can be simplified by
introducing a recursive procedure to obtain the $\left( n + 1
\right)$--point function from the $n$--point function.

The $n$--point function for the grand--canonical ensemble is written as
\begin{equation}
  \braket{\Psi(z) | \sigma^3_{x_1} \dots \sigma^3_{x_n} | \Psi(z)} = \frac{1}{2^n}\prod_{k=1}^n \frac{1-w\, \zeta_k}{1+w\,\zeta_k} (-w\, q^{1/2}; q)_\infty = f_w^{(n)} (\zeta_1, \dots, \zeta_n)  (-w\, q^{1/2}; q)_\infty \, , 
\end{equation}
where $w= z^2$ and $\zeta_i = q^{x_i + 1/2}$. It is convenient to
develop it in a series in $w$:
\begin{equation}
  f_w^{(n)} (\zeta_1, \dots, \zeta_n) = \sum_{k=0}^\infty w^k \lambda^{(n)}_k (\zeta_1, \dots,  \zeta_n)  \, .
\end{equation}
Passing to the $\left(n+1 \right)$--point function just amounts to
multiplying by an extra factor of $\frac{1-w\, \zeta_{n+1}}{1+w\,
  \zeta_{n+1}}$:
\begin{equation}
  2 f_w^{(n+1)} (\zeta_1, \dots, \zeta_n, \zeta_{n+1} ) = \frac{1-w\, \zeta_{n+1}}{1+w\, \zeta_{n+1}}  f_w^{(n)} (\zeta_1, \dots, \zeta_n) \, .
\end{equation}
Developing in a series,
\begin{equation}
  \frac{1-w\, \zeta_{n+1}}{1+w\, \zeta_{n+1}} = -1 + 2 \sum_{k=0}^\infty \left( -1  \right)^k w^k \zeta_{n+1}^k \, , 
\end{equation}
one finds
\begin{equation}
  f_w^{(n+1)} (\zeta_1, \dots, \zeta_n, \zeta_{n+1} ) = \sum_{k=0}^\infty w^k \lambda^{(n+1)}_k (\zeta_1, \dots,  \zeta_{n+1}) \, , 
\end{equation}
where
\begin{equation}
  \lambda_k^{(n+1)} (\zeta_1, \dots, \zeta_{n+1} ) = - \frac{1}{2} \lambda_k^{(n)} (\zeta_1, \dots, \zeta_n) + \left( -1 \right)^k \zeta_{n+1}^k \sum_{j=0}^k \left( - 1 \right)^j \lambda^{(n)}_j (\zeta_1, \dots, \zeta_n) \zeta_{n+1}^{-j} \, ,  
\end{equation}
which gives recursively the $\left(n+1 \right)$--point function for
the canonical ensemble $\lambda_k^{(n+1)}$ in terms of the $n$--point
function.  The recursion relation can be solved using as initial
condition the one--point function found in
App.~\ref{sec:calculations}:
\begin{equation}
  \lambda_k^{(1)} (\zeta_1) =
  \begin{cases}
    -\frac{1}{2} & \text{if $k=0$,} \\
    \left( - \zeta_1 \right)^k & \text{otherwise.}
  \end{cases}
\end{equation}
and yields
\begin{equation}
  \lambda_k^{(n)} (\zeta_1, \dots, \zeta_n) =
  \begin{cases}
    \left( -\frac{1}{2} \right)^n & \text{if $k=0$,} \\
    \displaystyle{\frac{1}{2^{n-1}} \sum_{i=1}^n} \left(\prod_{j=1}^{n-1} \frac{\zeta_i + \zeta_{i+j}}{\zeta_i - \zeta_{i+j}} \right) \left( - \zeta_i \right)^k & \text{otherwise,}
  \end{cases}
\end{equation}
where the indices $i$ and $j$ are understood in $\setZ_n$.

Now we can again use the same procedure outlined in
App.~\ref{sec:calculations} and find that on the $N$--particle kink,
\begin{equation}
  \braket{\sigma^3_{x_1} \dots \sigma^3_{x_n} }_N =  \frac{1}{\left(-2 \right)^n} + \frac{1}{2^{n-1}} \sum_{k=0}^N (q^{-N};q)_k \,q^{N k}\sum_{i=1}^n \left(\prod_{j=1}^{n-1} \frac{q^{x_i} + q^{x_{i+j}}}{q^{x_i} - q^{x_{i+j}}} \right)  q^{k x_i} \, ,
\end{equation}
and in the $N \to \infty $ limit this becomes:
\begin{equation}
  \braket{\sigma^3_{x_1} \dots \sigma^3_{x_n} }_\infty = \frac{1}{\left(-2 \right)^n} + \frac{1}{2^{n-1}} \sum_{j=0}^\infty \left( -1 \right)^j q^{\binom{j}{2}}  \sum_{i=1}^n \left(\prod_{j=1}^{n-1}  \frac{q^{x_i} + q^{x_{i+j}}}{q^{x_i} - q^{x_{i+j}}}  \right) q^{j x_i} \, .
\end{equation}
In the $q\to 1^- $ limit, this is
\begin{multline}
  \braket{\sigma^3_{x_1} \dots \sigma^3_{x_n} }_\infty =
  \frac{1}{\left(-2 \right)^n} - \frac{1}{2^{n-1}} \sum_{i=1}^n
  \left(\prod_{j=1}^{n-1} \coth (\frac{x_{i} - x_{i+j}}{2}) \right)
  \frac{e^{x_i}}{1+e^{x_i}} = \\ = \frac{1}{\left( - 2 \right)^n}
  \prod_{i=1}^n \tanh (\frac{x_i}{2}) = \prod_{i=1}^n
  \braket{\sigma^3_{x_i}}_\infty \, .
\end{multline}
We see that the system decorrelates completely.

\subsection{Height function and numerics for higher eigenstates}
\label{sec:numerics2d}

We would like to form an idea about the excited eigenstates of the 2d
quantum crystal. Since at this point, we do not have an exact
expression, we resort to numerical calculations for finite systems
containing partitions up to a certain number.

The height function is defined on the endpoints of the squares making
up the partitions. For a system allowing partitions fitting in a box
of length $N_0$, the height function is represented on a 1d integer
lattice of length $2 N_0 + 1$. We can visually represent any
eigenstate by plotting its average height. So we first have to express
each individual partition in terms of a height function and for a
given eigenstate sum all these heights with the probability
coefficients of this eigenstate.  More precisely, after the
quantization of the system the height function becomes an observable
and as such it has an expectation value for a given state
$\ket{\Psi}$,
\begin{equation}
  \braket{h(\mathbf{n})}_\Psi = \frac{\braket{\Psi| h (\mathbf{n}) | \Psi}}{\braket{\Psi | \Psi} } = \frac{1}{\norm{\Psi}^2} \sum_{\alpha} h_{\alpha} (\mathbf{n}) \abs{\braket{\Psi|\alpha }}^2 = \sum_{\alpha} h_{\alpha}  (\mathbf{n}) m_{\Psi} ( \alpha ) \, ,
\end{equation}
where the sum extends over the partitions $\ket{\alpha}$.

We consider the contribution of each partition to the height function
minus the profile of the empty corner, which has the form
$\set{\dots,\, 2,\, 1,\, 0,\, 1,\, 2,\, \dots}$. A partition of $N$ is
given by $N$ positive integers $(a_0,\, a_1, \dots, a_{N-1})$ with
$\sum_{i=0}^{N-1} a_i=N$. Each $a_i$ gives rise to a vector
$\delta_{a_i}$ with all entries zero except for a sequence of $a_i$
twos from position $i-a_i+1$ to $i$. The sum of these $N$
contributions gives the profile of the partition. The height function
of a partition $\mu$ at position $m$ is thus given by
\begin{equation}
  h_\mu(m) = h_0(m) + 2 \sum_{i=0}^{N-1}\chi_{[i-a_i+1,i]}(m) = h_0 (m) + \delta_\mu (m) \, ,
\end{equation}
where $h_0$ is the height function of the empty corner and
$\chi_{[a,b]}(m)$ is the characteristic function of the interval
$[a,b]$.  To make an example, the partition $(3,2,2)$ gives rise to
the profile
\begin{equation}
  h_{(3,2,2)} = \set{3,4,3,4,5,4,3}
\end{equation}
consisting of the profile of the empty partition $h_{(0,0,0)}= \set{3,2,1,0,1,2,3}$ and 
\begin{equation}
  \begin{split}
    \delta_{(3,2,2)} &= \set{0,2,2,4,4,2,0} \\
    &= \set{0,2,2,2,0,0,0}\\
    &+ \set{0,0,0,2,2,0,0}\\
    &+ \set{0,0,0,0,2,2,0}.
  \end{split}
\end{equation}

\begin{figure}
  \begin{center}
    \includegraphics[width=.8\textwidth]{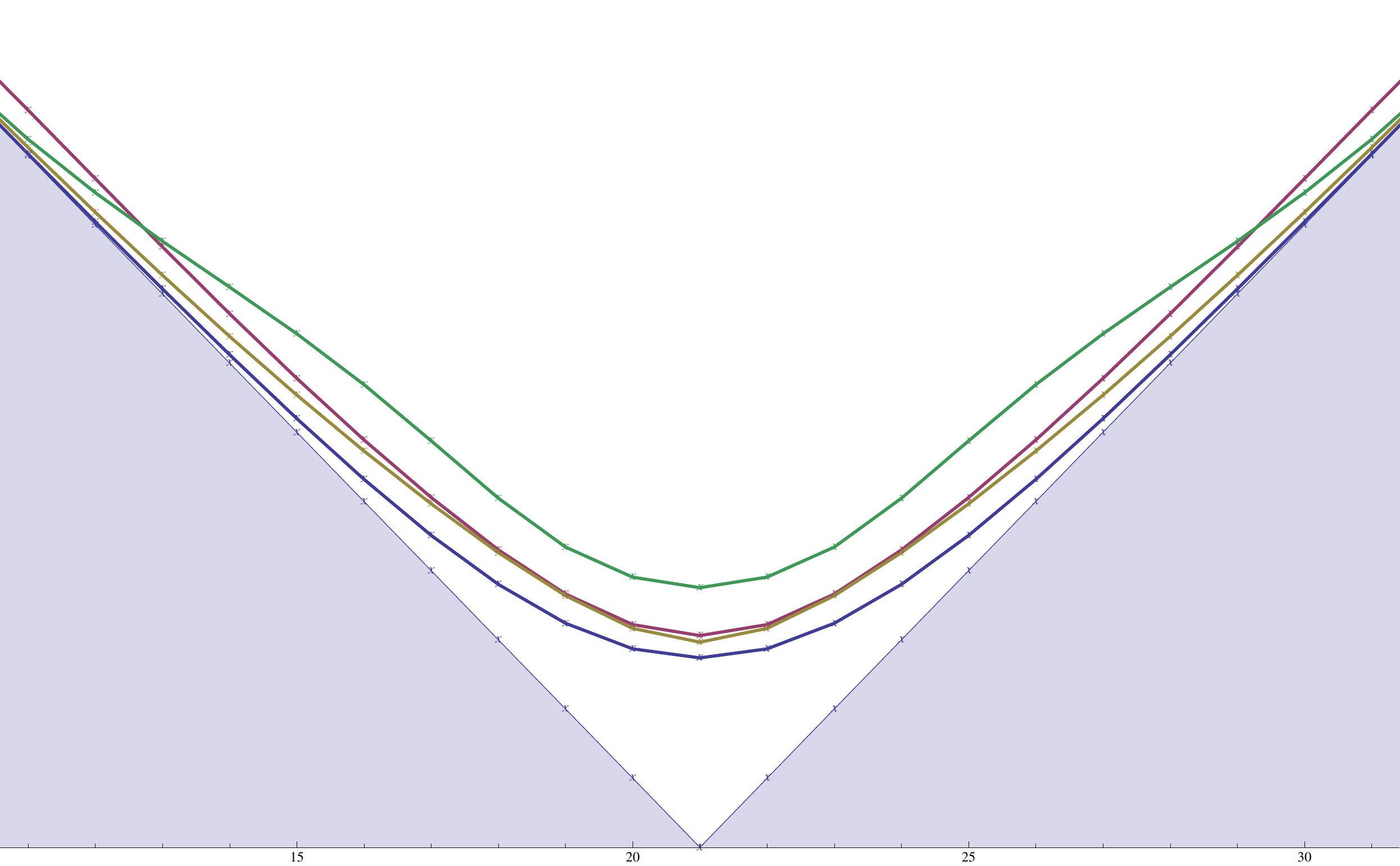}  
    \caption{Average heights for eigenstates of a system of partitions
      of up to $20$ (corresponding to a total of $2714$ partitions) and
      $q=0.9$. Ground state (blue), first excited state (yellow),
      fourth excited state (red), twelfth excited state (green).}
    \label{fig:numeric2de}
  \end{center}
\end{figure}

The average heights (\ie the integrals of the average magnetization of
the underlying spin chains) of the ground state, first, fourth and
twelfth eigenstates of a system containing partitions of integers up
to 20 (which corresponds to 2714 partitions) for a value of $q=0.9$
are depicted in Figure~\ref{fig:numeric2de}.

The ground state corresponds to the curve we have calculated
explicitly in Sec.~\ref{sec:one-point-function}, see
Figure~\ref{fig:magnetization-limitshape}(b). The profiles of the next
excited states are a bit flatter. Higher excited states such as the
twelfth shown here can show a deviation from this pattern.


%% file: ThreeDimensions.tex
\section{The 3d Hamiltonian}
\label{sec:3d-hamiltonian}

\subsection{Fermionic formalism}
\label{sec:fermionic-formalism}

The general problem of the 3d melting crystal corner has already been
explained in Section~\ref{sec:qcrsytalmelting}. We now want to express
the 3d quantum Hamiltonian in terms of fermionic operators,
analogously to the treatment of the 2d case in
Section~\ref{sec:fermion2d}.

\begin{figure}
  \begin{center}
    \subfigure[Cubes]{\includegraphics[width=.25\textwidth]{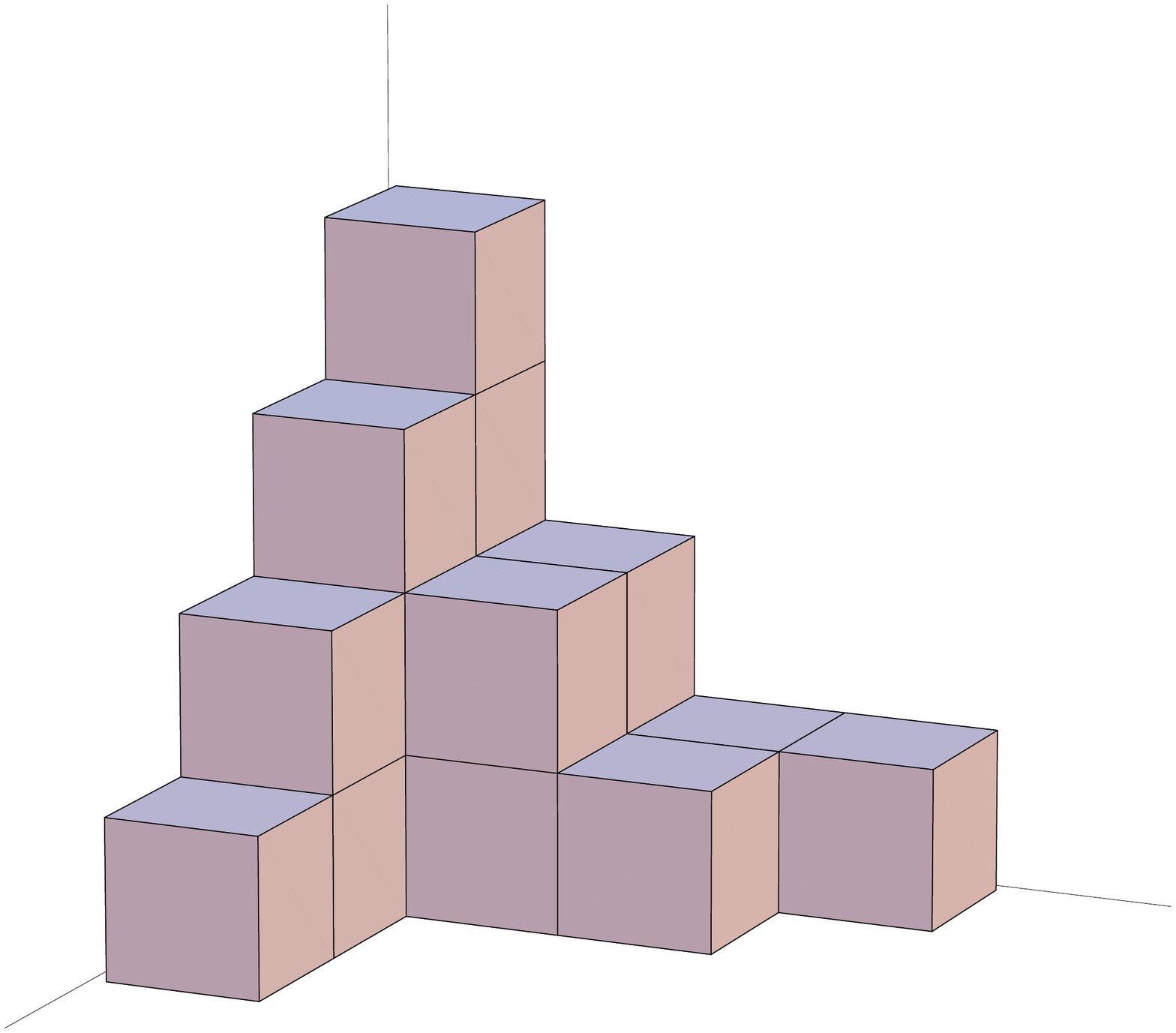}} \hfill
    \subfigure[Vicious walkers]{\includegraphics[width=.25\textwidth]{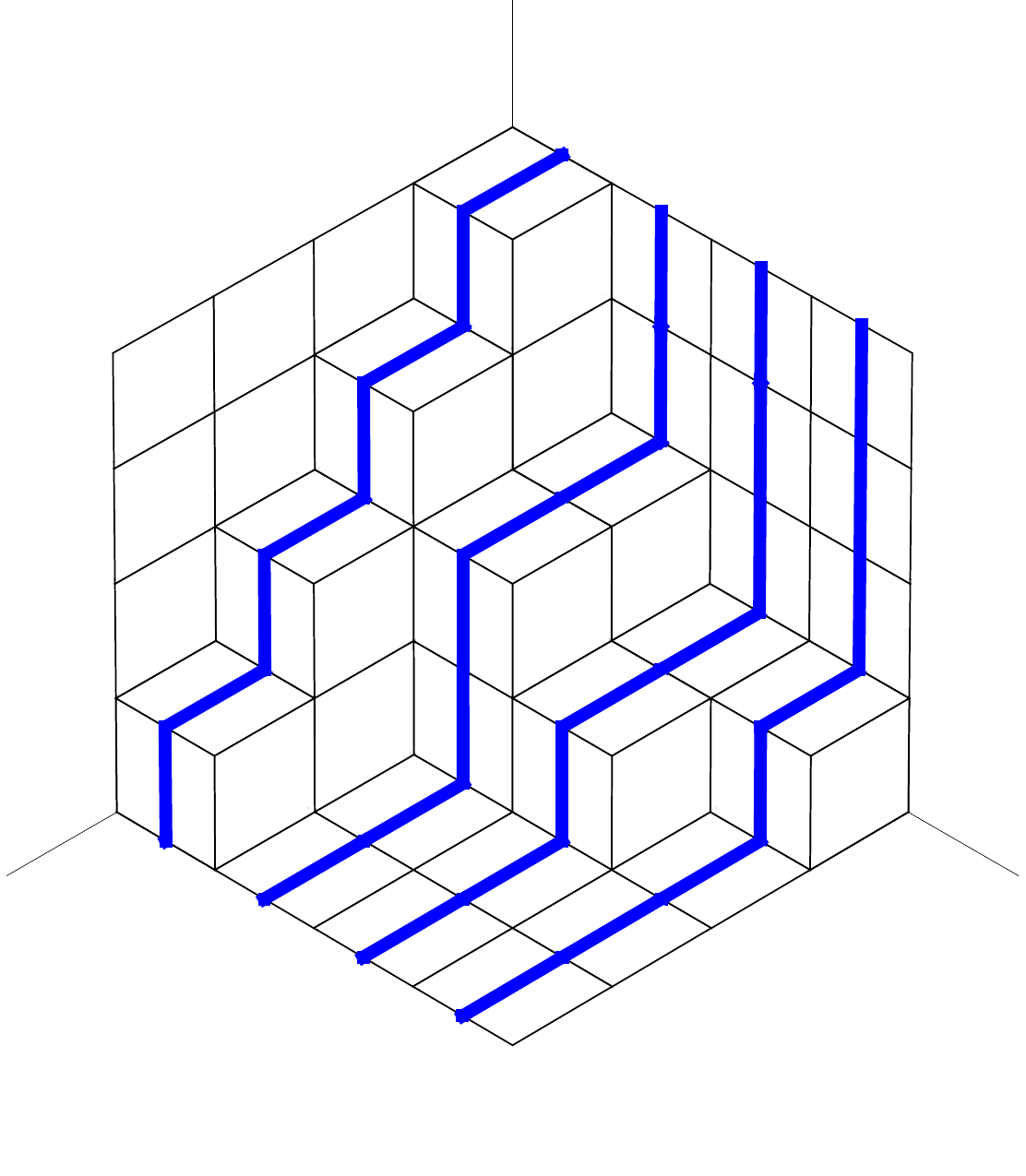}} \hfill 
    \subfigure[Vicious walkers on the square lattice]{\includegraphics[width=.18\textwidth]{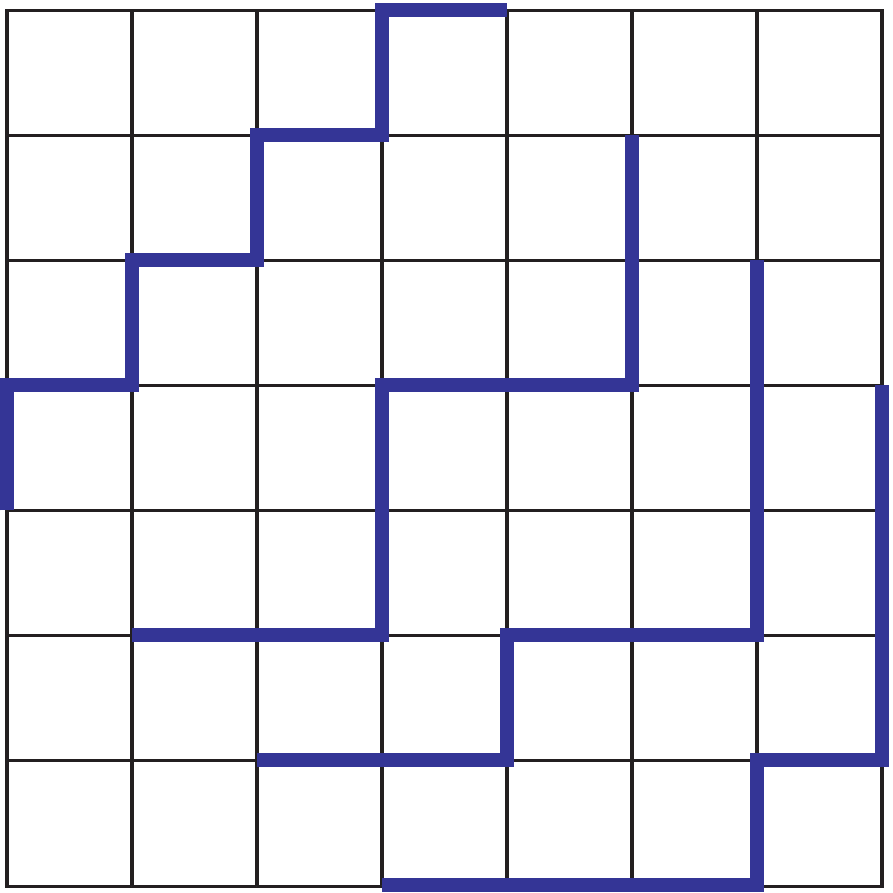}} \hfill
    \subfigure[2d fermions]{\includegraphics[width=.18\textwidth]{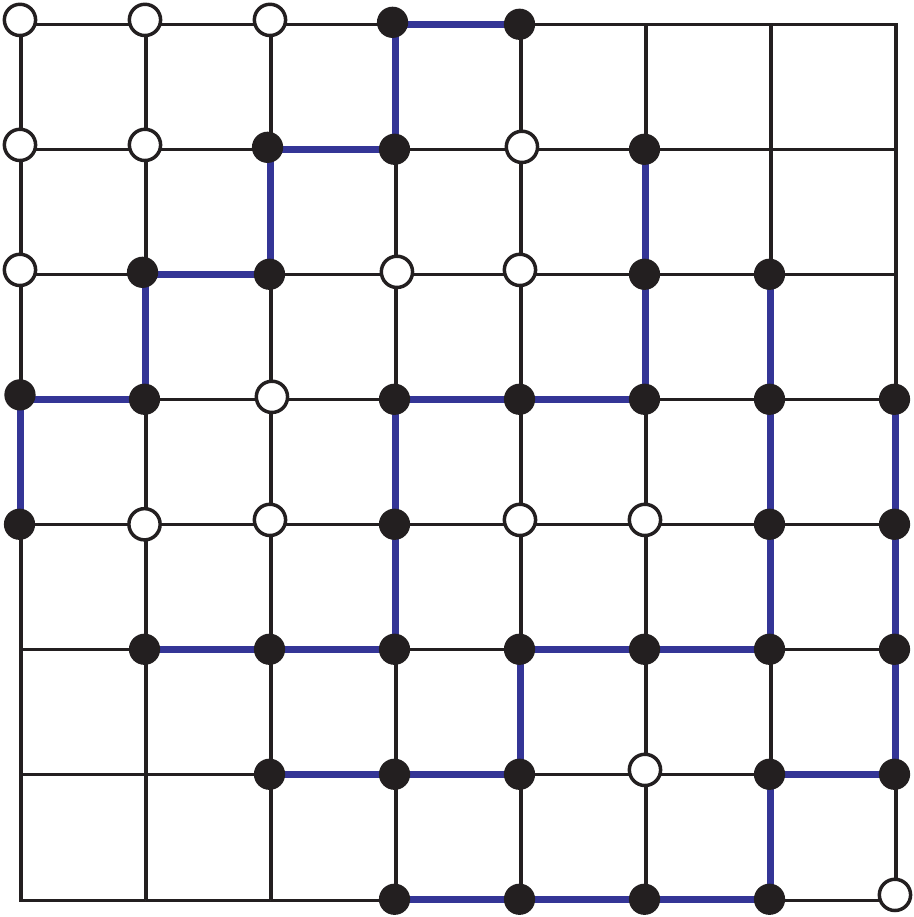}} 
  \end{center}
  \caption{Vicious walkers and three--dimensional partitions. Representation of a three--dimensional partition as cubes (a); as a system of vicious walkers (b); as vicious walkers in a square lattice (c) and as a system of 2d fermions (d).}
  \label{fig:3d-vicious}
\end{figure}

Not surprisingly, the quantum Hamiltonian for the three--dimensional
partitions can be recast in terms of a system of Heisenberg
\textsc{xxz} chains. To do so, we first map the three--dimensional
partitions into a system of vicious walkers. Vicious walkers were
introduced by Fisher in~\cite{Fisher2} as a system of $N$ particles
moving in non--intersecting random walks. The mapping to plane
partitions is obtained as follows. Consider the finite problem in a
cubic box of length $N$.  The empty partition can be represented as an
equilateral hexagon tessellated with rhombi.  Choose the midpoints of
the sides of these rhombi on the lower left side of the hexagon and
join them with the corresponding midpoints of the rhombi on the upper
right edge, following the sides of the cubes
(Fig.~\ref{fig:3d-vicious}(b)). One can verify that each partition
corresponds to a realization of $N$ non--intersecting directed paths,
\emph{i.e.} directed vicious walks on the lattice obtained by taking
the midpoints of the rhombi in the tessellation of the empty
partition. In particular, the empty partition is obtained as a set of
parallel walks that first move to the right and then up.

\begin{figure}
  \begin{center}
    \subfigure[Cubes]{\includegraphics[width=.3\textwidth]{Simple3dCubes}} 
    \subfigure[Diagonal slicing]{\includegraphics[width=.3\textwidth]{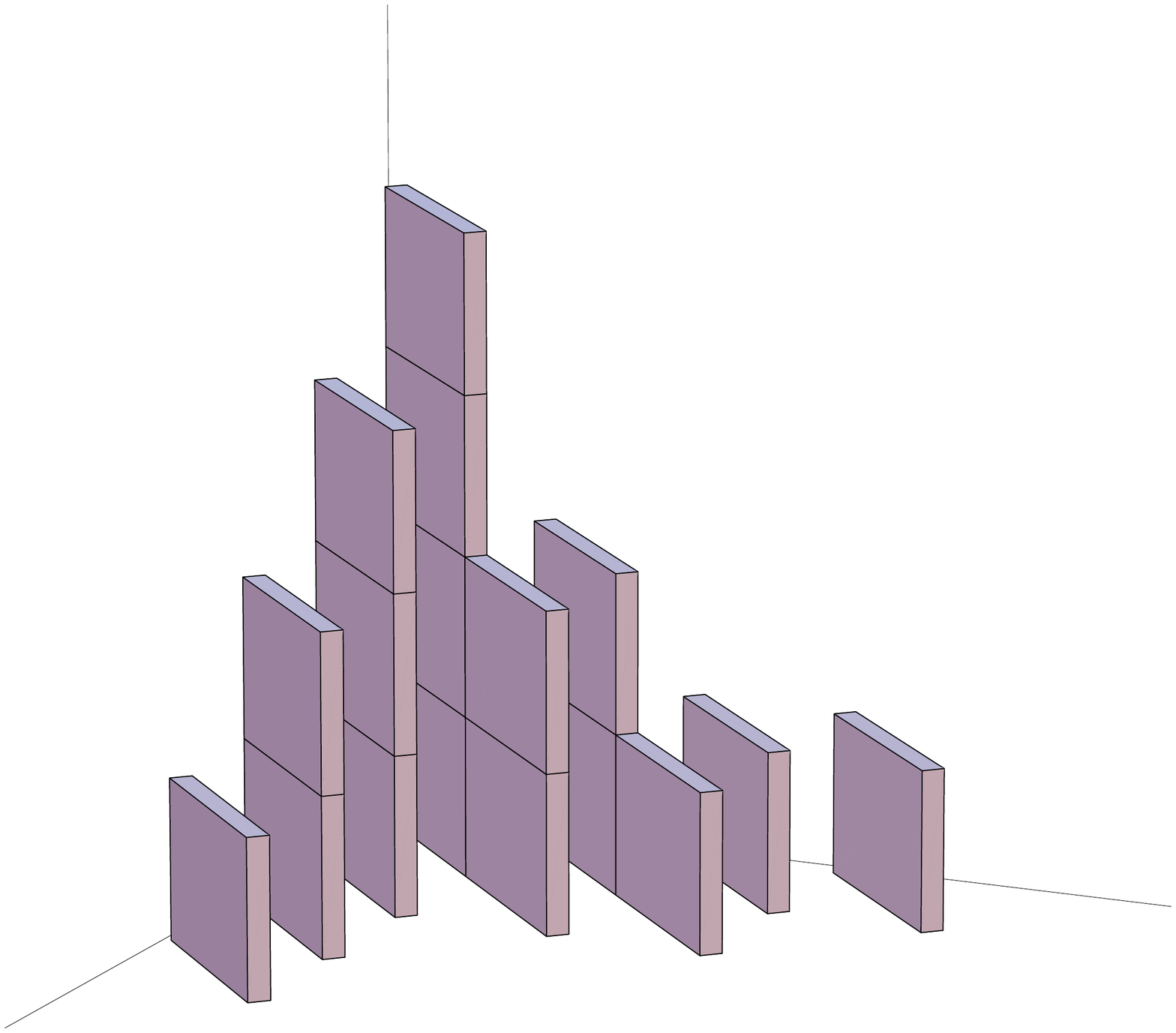}}
    \subfigure[Lattice fermions]{\includegraphics[width=.3\textwidth]{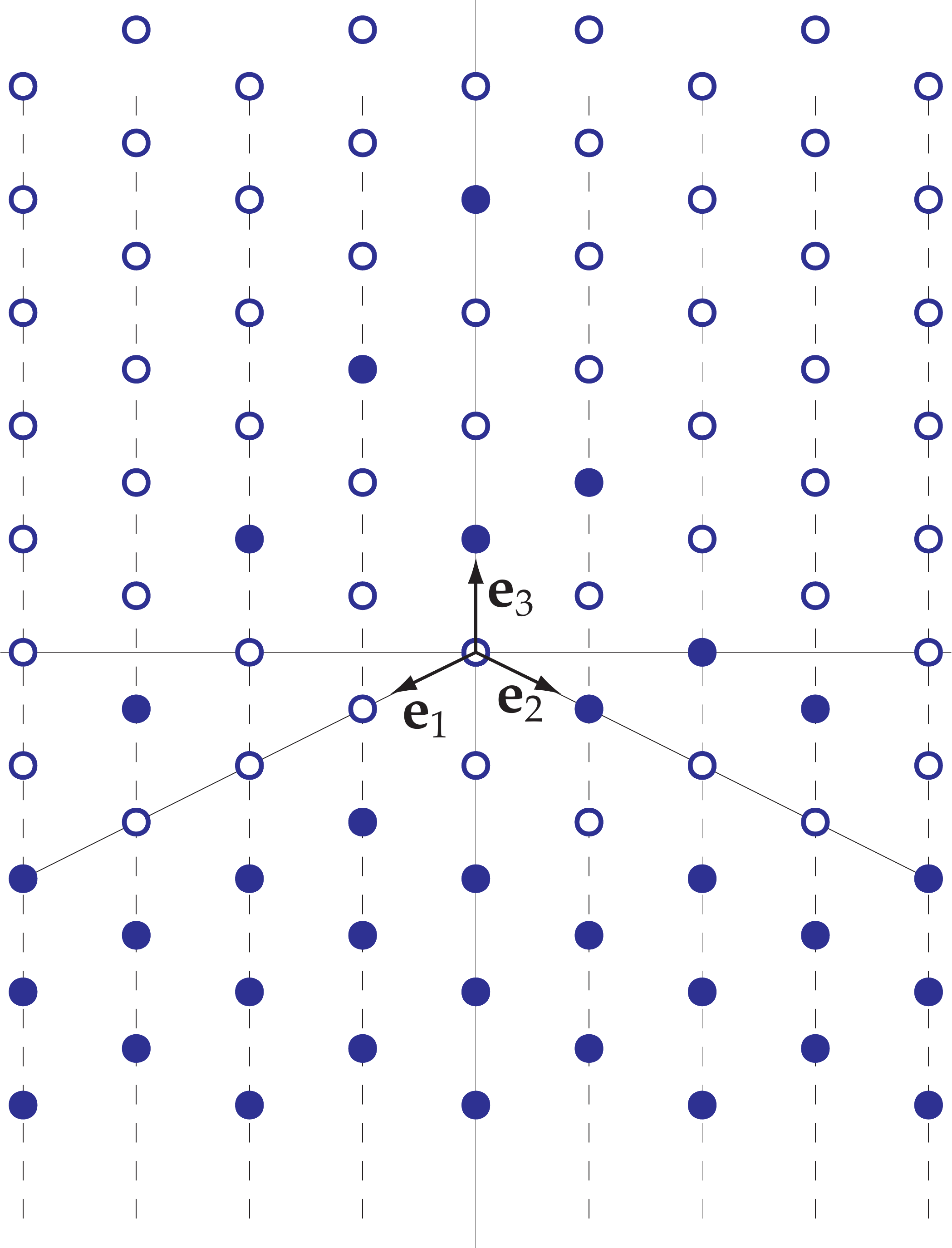}} 
  \end{center}
  \caption{Representation of a three dimensional partition as cubes
    (a); slicing into 2d partitions (b); lattice fermions (c). In the
    empty partition case the fermions occupy all the points below the
    diagonal lines.}
  \label{fig:3d-slicing-fermions}
\end{figure}

It is convenient to represent the system on a square lattice by
tilting the picture (Fig.~\ref{fig:3d-vicious}(c)). One can easily see
that the shape of each vicious walk corresponds to a Young diagram
obtained by parallel--slicing the three--dimensional partition. The
set of $N$ paths can be further mapped to a two--dimensional system of
fermions living on the vertices of the lattice joined by a path
(Fig.~\ref{fig:3d-vicious}(d)).  Let us now consider the quantum
Hamiltonian of this system. Adding a cube corresponds to moving a
fermion from $(i,j)$ to $(i-1,j+1)$. But this can only happen if the
points $(i-1,j)$ and $(i,j+1)$ are already occupied. Similarly,
removing a cube corresponds to jumping in the opposite way, respecting
the same conditions. The Hamiltonian can therefore be written as
\begin{equation}
  \label{eq:3d-Hamiltonian-vicious}
  \HH_{\text{3d}}  = -J \sum_{i,j} \HH_{\text{2d}}^{(i,j) \to (i-1, j+1)} n_{i-1,j} n_{i,j+1} \, ,
\end{equation}
where
\begin{equation}
  \HH_{\text{2d}}^{(i,j) \to (k, l)} = \psi^*_{k,l} \psi_{i,j} + \psi^*_{i,j} \psi_{k,l} - q^{1/2} n_{i,j} \left( 1 - n_{k,l} \right) - q^{-1/2} n_{k,l} \left( 1 - n_{i,j} \right) 
\end{equation}
is the Hamiltonian of the two--dimensional problem\footnote{In the
  study of the three--dimensional system we restrict ourselves to the
  \textsc{rk} point $J=V$. Note that while it is easy to generalize
  the lower--dimensional models (as already commented after
  Eq.~\ref{eq:Heisenberg}), in the case at hand the parameter $V$
  cannot be reabsorbed by redefining $\Delta$.}. We find that the
Hamiltonian describes an infinite system of parallel \textsc{xxz}
chains, each interacting with the two neighbouring ones. Equivalently,
the system also represents the dynamics of a heap of
dimers~\cite{Viennot:1985,Viennot:1986}. Interestingly enough, the
same Hamiltonian is obtained when considering a diagonal slicing as
shown in Fig.~\ref{fig:3d-slicing-fermions} instead of the vicious
walker description which amounts to a parallel slicing\footnote{In
  this diagonal slicing description, the black dots in
  Fig.~\ref{fig:3d-slicing-fermions}.(c) are allowed to move up if not
  only the next position on the same vertical line is free, but also
  the next positions above on both neighbouring spin chains are
  unoccupied.}.

There is a direct mapping between the rhombus tiling making up the
shape of the cubes (see Figure~\ref{fig:3d-vicious}(b)) and the dimer
model on the hexagonal graph (see also
Figure~\ref{fig:honey_cubes}). To the long diagonal of each rhombus we
associate an edge covered by a dimer in the underlying lattice. In
this way, a perfect matching determines the orientations of the
various rhombi in the tiling and therefore the shape of the 3d
partition. From the dimer picture on the hexagonal lattice we can
arrive directly at the fermion picture in
Figure~\ref{fig:3d-slicing-fermions}(c) by drawing a black dot for
each horizontal line covered by a dimer and drawing a white dot for
each uncovered horizontal line on the hexagonal graph. This also leads
directly to the half--integer shift between adjacent spin
chains. Furthermore, the dependence on the two neighboring spin chains
becomes manifest, since information about the horizontal lines of the
neighboring cells is needed to reconstruct the complete perfect
matching from the knowledge of the dimers on the horizontal lines
only.

\subsection{Hexagonal lattice description}
\label{sec:hexag-latt-descr}

We can make use of the correspondence to the dimer model on the
hexagonal lattice and describe the fermions in the plane more
conveniently using an overcomplete basis.  The plaquettes of a planar
hexagonal lattice are labeled by a triplet of positive integers
$\set{n_1, n_2, n_3}$ such that the center of each hexagon has
coordinates $n_1 \mathbf{e}_1 + n_2 \mathbf{e}_2 + n_3 \mathbf{e}_3
$. The $\mathbf{e}_i$ are unit vectors that can be chosen to be
\begin{align}
  \mathbf{e}_1 = ( - \cos
\tfrac{\pi}{6}, - \sin \tfrac{\pi}{6}) \, , && \mathbf{e}_2 = ( \cos
\tfrac{\pi}{6}, - \sin \tfrac{\pi}{6})\,, && \, \text{and} && \mathbf{e}_3 = (0,1) \, .
\end{align}
Note that the three unit vectors are not linearly independent and
satisfy $\mathbf{e}_1 + \mathbf{e}_2 + \mathbf{e}_3 = 0$ which implies
that two triplets $\set{n_1, n_2, n_3}$ and $\set{m_1, m_2, m_3 }$
represent the same point if $\sum_{i=1}^3 \left( m_i - n_i \right)
\mathbf{e}_i = 0$.  We thus obtain a system where each fermion can
jump from the point $\mathbf{x}$ to $\mathbf{x} + \mathbf{e}_3$ if
both the points $\mathbf{x} - \mathbf{e}_1$ and $\mathbf{x} -
\mathbf{e}_2$ are free (see
Fig.~\ref{fig:3d-slicing-fermions}(c)). Accordingly, we rewrite the
Hamiltonian in Eq.~(\ref{eq:3d-Hamiltonian-vicious}) as
\begin{equation}
  \label{eq:3d-Hamiltonian}
  \HH_{\text{3d}}  = -J \sum_{\mathbf{x}} \HH_{\text{2d}}^{\mathbf{x} \to \mathbf{x} + \mathbf{e}_3} \left( 1- n_{\mathbf{x} - \mathbf{e}_1} \right) \left( 1 -  n_{\mathbf{x} - \mathbf{e}_2} \right) \, .
\end{equation}
In this notation, the empty partition corresponds to the state
\begin{equation}
  \ket{\text{half}} = \prod_{n_1, n_2 = 0}^\infty \psi^*_{n_1 \mathbf{e}_1 + n_2 \mathbf{e}_2} \ket{0} \, .    
\end{equation}
A generic partition is expressed by
\begin{equation}
  \label{eq:generic-3d-partition}
  \ket{\mu } = \prod_i \psi^*_{\mathbf{x}_i  + \gamma_i \mathbf{e}_3} \psi_{\mathbf{x}_i}  \ket{\text{half}} \, ,
\end{equation}
where $\mathbf{x}_i = \alpha_i \mathbf{e}_1 + \beta_i \mathbf{e}_2 $
and $\alpha_i $, $\beta_i $, and $\gamma_i$ are all positive. To be
admissible, a set $\set{\mathbf{x}_i, \gamma_i}$ has to satisfy
certain interlacing conditions to make sure it defines a valid plane
partition. These conditions are particularly simple in this notation
once one realizes that $( \alpha_i, \beta_i , \gamma_i)$ are the
coordinates of one of the corners of the topmost box in each column
forming the three--dimensional partition. This means that for each
column we have to verify that the two columns behind are taller,
namely:
\begin{equation}
\label{eq:interlacing-spins}
  \gamma_i \leq \gamma_j \, \text{ if $\mathbf{x}_i = \mathbf{x}_j + \mathbf{e}_1$ or $\mathbf{x}_i = \mathbf{x}_j + \mathbf{e}_2$}.
\end{equation}
The number of boxes in a given partition $\mu$ is
\begin{equation}
  \abs{\mu } = \sum_i \gamma_i \, .
\end{equation}
The ground state of the quantum crystal expressed in fermionic
language is given by
\begin{equation}
  \label{eq:ground3dfer}
  \ket{\text{ground}} =\left[\sum_{\set{\mathbf{x}_i,\gamma_i}} q^{\frac{1}{2}\sum_i \gamma_i} \prod_i \psi^*_{\mathbf{x}_i  + \gamma_i \mathbf{e}_3} \psi_{\mathbf{x}_i}  \right] \ket{\text{half}} \, ,  
\end{equation}
where the sum runs over all the sets $\set{\mathbf{x}_i,\gamma_i}$ satisfying the condition given in Eq.~(\ref{eq:interlacing-spins}).

\bigskip

This representation permits an explicit mapping between
three--dimensional partitions and perfect matchings on the (bipartite) hexagonal
lattice. Each perfect matching $PM$ defines a unique height function
on the plaquettes, up to a constant. This function $h_{PM}$ is defined
as follows:
\begin{equation}
  h_{PM} (A) - h_{PM} (B) =
  \begin{cases}
    \pm 1 & \text{if $e \in PM$,} \\
    \mp 2 & \text{if $e \not \in PM$,} \\
  \end{cases}
\end{equation}
where $A $ and $B$ are two plaquettes that share the edge $e$, the
sign depends on the orientation\footnote{Given the bipartite nature of
  the hexagonal lattice, we can divide vertices in black and
  white. The sign depends on whether $e$ is crossed from $A$ to $B$
  with a black or a white node on the left.}. The empty room height
function corresponds to the unique (up to translation) empty room
perfect matching (See Figure~\ref{fig:honey_cubes}.(a)) and is given
by
\begin{equation}
  h_0 ( n_1 \mathbf{e}_1 + n_2 \mathbf{e}_2 + n_3 \mathbf{e}_3 ) =  n_1 + n_2 + n_3 - 3 \min \set{n_1, n_2, n_3} + k   \, ,  
\end{equation}
where $k$ is a constant which we set here to $k=0$.  

The map from the partition $\mu$ expressed as in
Eq.~(\ref{eq:generic-3d-partition}) to the height function $h^\mu
(\mathbf{y})$ and hence a perfect matching is obtained as follows:
\begin{equation}
  h_\mu ( \mathbf{y} ) = h_0 ( \mathbf{y} ) + 3 \sum_i \sum_{n=0}^{\gamma_i - 1} \delta ( \mathbf{y} - \mathbf{x}_i - n \mathbf{e}_3 ) = h_0 (\mathbf{y}) + \delta_\mu (\mathbf{y}) \, ,
\end{equation}
where
\begin{equation}
  \delta (\mathbf{x}) =
  \begin{cases}
    1 & \text{if $\mathbf{x} = 0$,} \\
    0 & \text{otherwise.}
  \end{cases}
\end{equation}
Using these normalizations,
\begin{equation}
  \abs{\mu} = \frac{1}{3} \sum_{\mathbf{y}} \delta_\mu (\mathbf{y}) \, .  
\end{equation}

\subsubsection*{Numerical results for higher eigenstates}

\begin{figure}[p]
  \begin{center}
    \subfigure[Ground state]{\includegraphics[width=.2\textwidth]{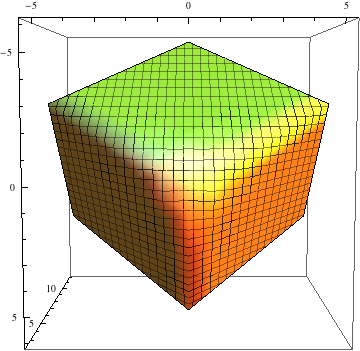}}
    \hfill
    \subfigure[First excited]{\includegraphics[width=.2\textwidth]{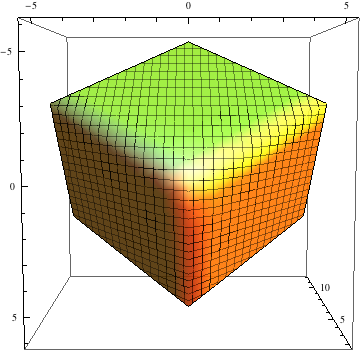}}
    \hfill
    \subfigure[Second excited]{\includegraphics[width=.2\textwidth]{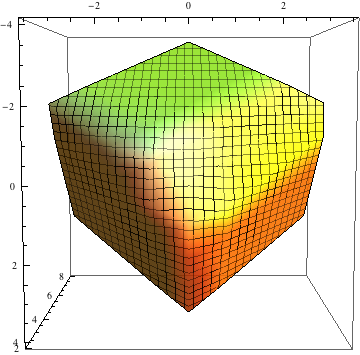}}
    \hfill
    \subfigure[Second excited (degenerate)]{\includegraphics[width=.2\textwidth]{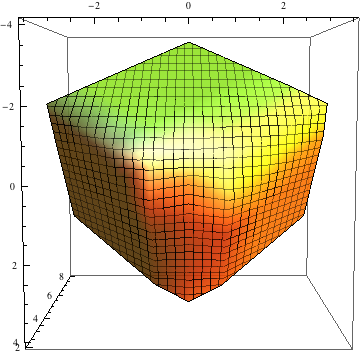}}
    \hfill
    \subfigure[Ground state]{\includegraphics[width=.2\textwidth]{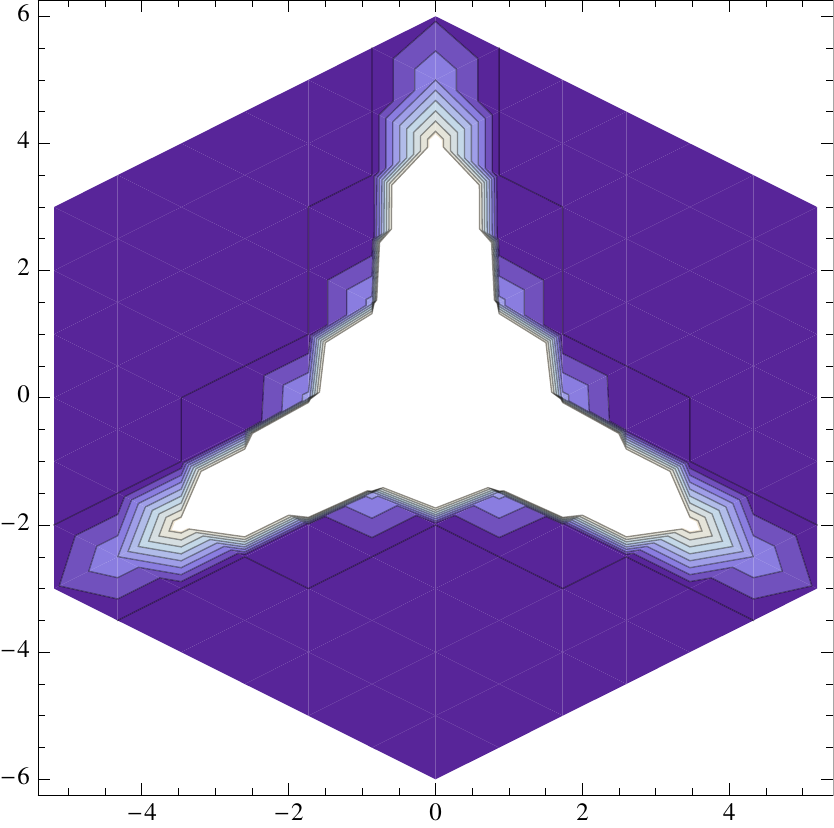}}
    \hfill
    \subfigure[First excited]{\includegraphics[width=.2\textwidth]{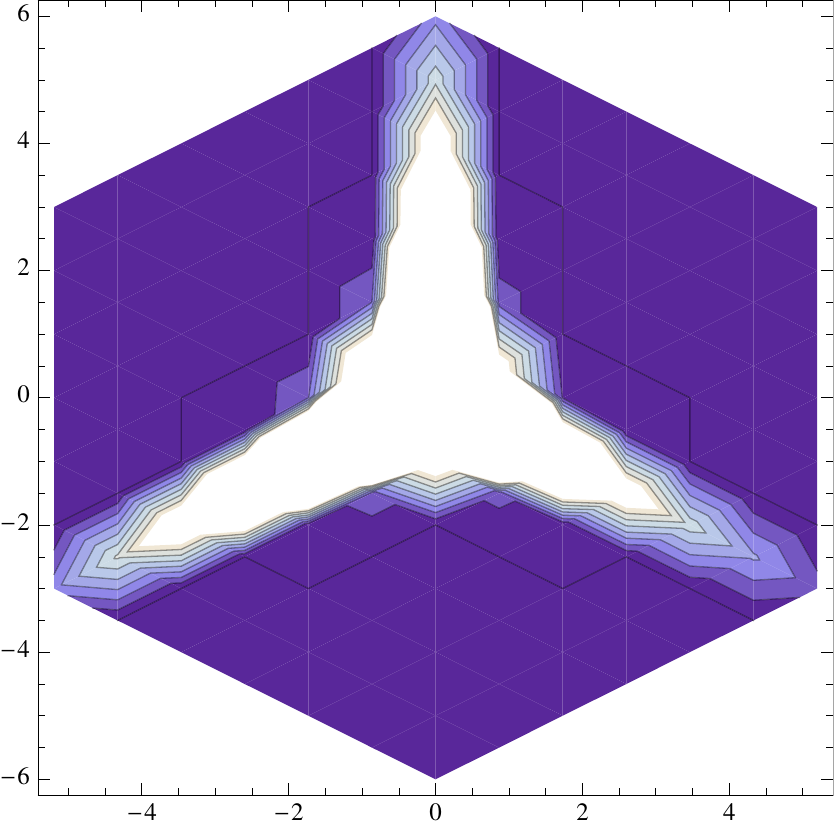}}
    \hfill
    \subfigure[Second excited]{\includegraphics[width=.2\textwidth]{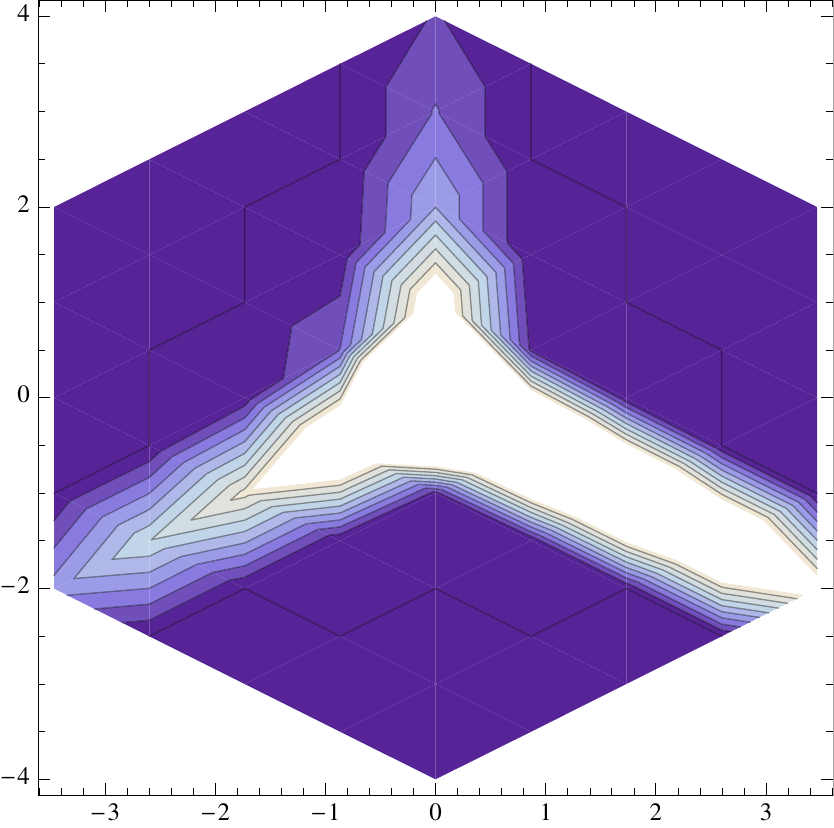}}
    \hfill
    \subfigure[Second excited (degenerate)]{\includegraphics[width=.2\textwidth]{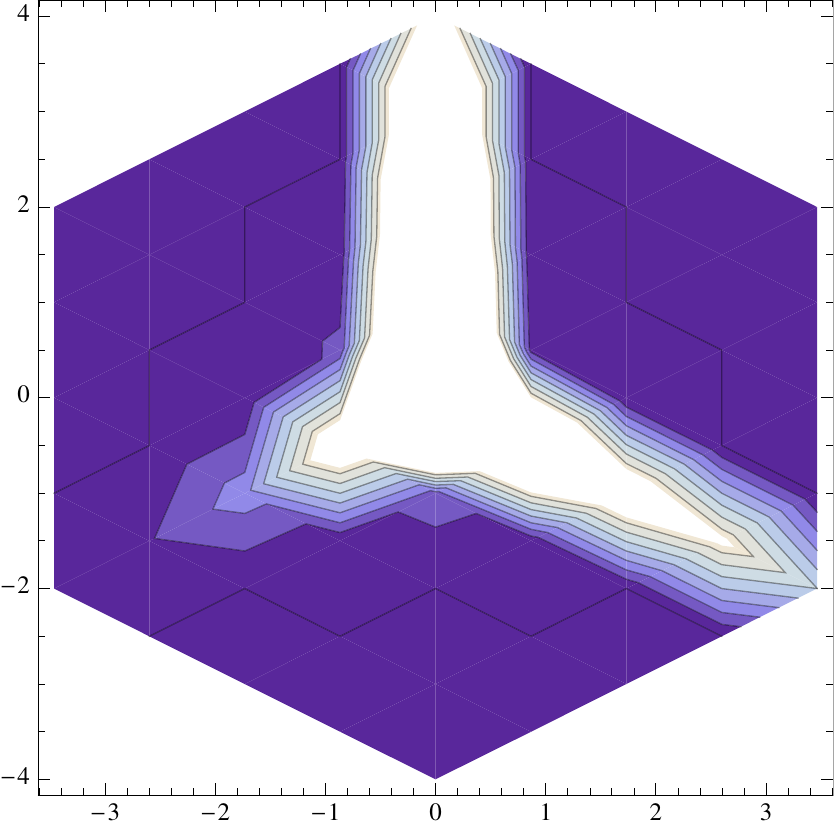}}
    \hfill
  \end{center}
  \caption{Ground state and first 3 eigenvectors of the three
    dimensional quantum crystal for plane partitions up to $12$ (in total
    of $3462$ configurations) and $q=0.8$. Molten corner and
    contour plot of the height function.}
  \label{fig:3d-Eigenvectors}
\end{figure}

As already explained in Sec.~\ref{sec:numerics2d}, we are interested in
the average height function for the excited states. An
analytic expression is at this point beyond our reach. We have
nevertheless run some numeric simulations for a system containing
plane partitions of integers of up to $12$ (resulting in a total of
$3462$ three--dimensional configurations). Some of the results are shown in
Figure~\ref{fig:3d-Eigenvectors}. In particular one finds that the
excited levels are more spread out than the ground state and we have
some evidence that the first excited level has $\setZ_3$ symmetry and
is non--degenerate, while the second excited level has only $\setZ_2$
symmetry and is thrice degenerate.

\subsection{The mass gap}
\label{sec:mass-gap}

\begin{figure}[p]
  \begin{center}
    \includegraphics[width=.75\textwidth]{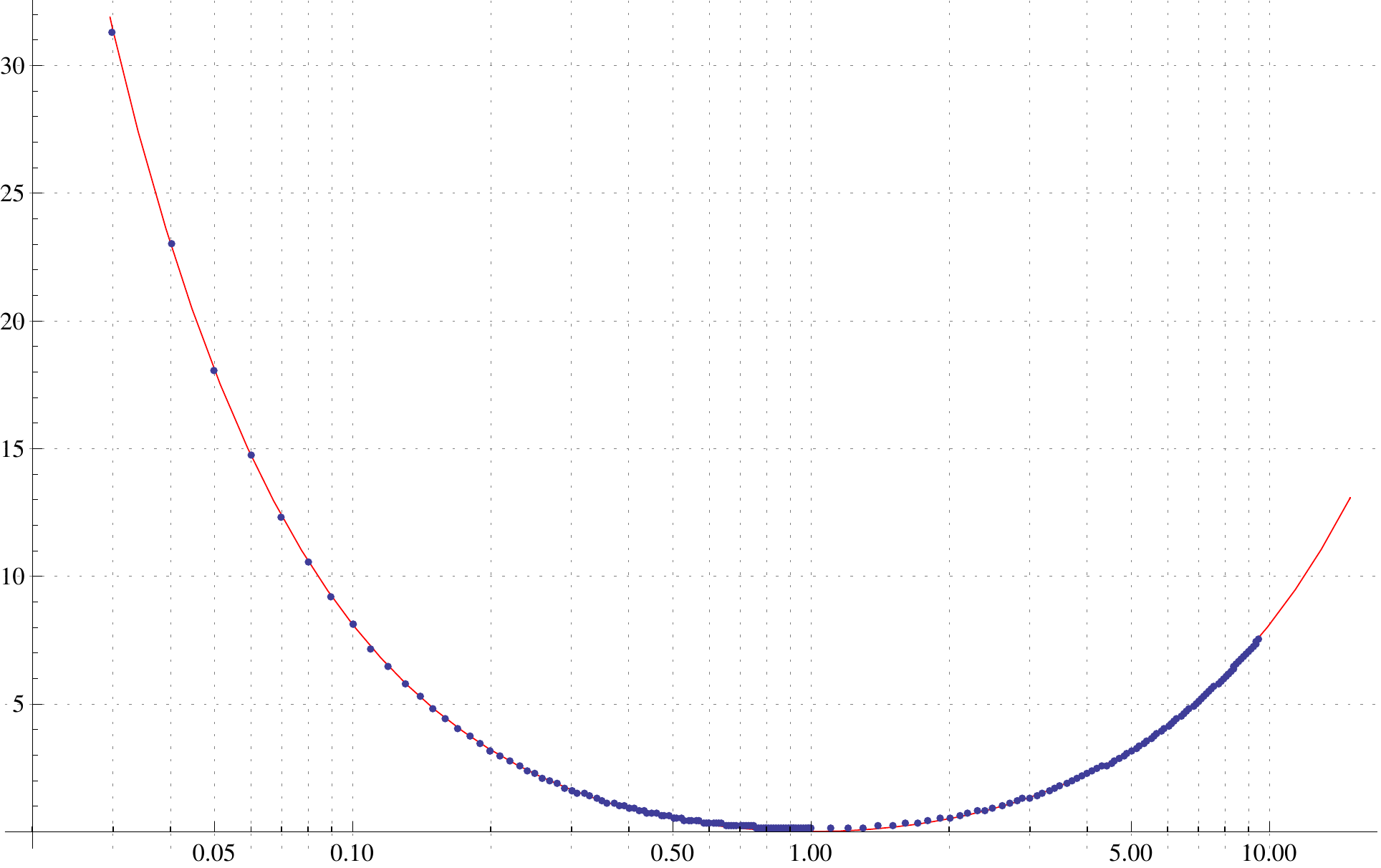}
  \end{center}
  \caption{The mass gap as a function of $\sqrt{q}$. The red line corresponds to
    $y = q^{1/2} + q^{-1/2} - 2$. The blue points result from a
    numerical simulation in a cube of size $3 \times 3 \times 4$ with a
    total of 4116 explored three--dimensional partitions.}
  \label{fig:gap-3d}
\end{figure}

We have remarked already that the mass gaps of the 1--dimensional (see
formula~(\ref{eq:gap1d})) and two--dimensional quantum crystals (see
Eq.~(\ref{eq:gap})) are the same. For the 3d--quantum crystal, we have
strong numerical evidence for the mass gap taking again at the same
value, see Figure~\ref{fig:gap-3d}. We therefore conjecture that
\begin{center}\emph{The mass gap of the quantum crystal is the same in all
  dimensions and equals to} $\gamma=-J(2-\sqrt q-1/\sqrt q)$.
  \end{center}
Why this should be true can heuristically be understood as
follows. Consider the graph of configurations
(Fig.~\ref{fig:state-diagram}), which is a directed, acyclic
graph. This graph admits a height function, the height of a
configuration being its number of boxes. The quantum crystal melting
can be considered an asymmetric random walk on this graph, where a
particle hops from one configuration to the next. This random walk
corresponds directly to the 1--particle sector of an \textsc{xxz}
model living on the state graph. In one dimension, this corresponds
directly to the \textsc{xxz} model on the line with one particle
moving around. In two dimensions, it corresponds to the half--filled
sector of the 1d \textsc{xxz} model and in three dimensions to the
quantum crystal.  In all cases the Hamiltonian takes the form
\begin{equation}
  \HH = -J \left[ \triangle - \left( q^{1/2} - 1 \right) \deg^-(\alpha ) - \left( q^{-1/2} - 1 \right)  \deg^+ (\alpha ) \right]  \, ,  
\end{equation}
where $\deg^-(\alpha )$ and $\deg^+(\alpha )$ are the outdegree and
indegree of the node $\alpha $.  We would like to argue that the mass
gap is the same for the 1--particle sector of the \textsc{xxz} model
on any directed, acyclic graph satisfying the condition
\begin{equation}
  \frac{\deg^+(\alpha )}{\deg^-(\alpha )} \xrightarrow[\deg^-(\alpha )\to\infty]{} 1.
\end{equation}
This is the case in the one--dimensional system in
Sec.~\ref{sec:one-dimens-young} ($\deg^-(\alpha ) = \deg^+(\alpha ) =
1$) and the two--dimensional system of Sec.~\ref{sec:2d_partitions}
($\deg^- (\alpha ) = \deg^+(\alpha ) + 1$). We conjecture it to be the
same in three dimensions for the quantum crystal.

A similar result was already proven in~\cite{Handjani} for the case of
the graph of configurations being a tree.


%% file: Conclusions.tex
\section{Conclusions}
\label{sec:conclusions}

In this note we have studied systems of crystal melting in one, two
and three dimensions. Generalizing the work of Rokhsar and Kivelson on
the quantum dimer we obtain a quantum Hamiltonian whose unique ground
state reproduces the classical Boltzmann distribution for the
canonical ensemble.

The Hamiltonian becomes tractable once one realizes that it is
possible to introduce a mapping to spin systems. More precisely, we
found that the two--dimensional quantum crystal problem (which can also be
phrased in terms of random partitions) is \emph{integrable}, since it
corresponds directly to the half--full sector of the \textsc{xxz}
ferromagnetic spin chain. In particular, the random height function can
be interpreted as the integral of the magnetization of the underlying
spin system. In a similar fashion, the three--dimensional quantum crystal
can be described in terms of an infinite system of coupled
\textsc{xxz} spin chains. This, in turn, allowed us to carry out a
numerical analysis, leading to the conjecture of a mass gap
whose value does not depend on the dimensionality of the problem.

From this point on there is a variety of interesting questions to
address, mainly concerning the three dimensional system.  A rigorous
proof for the value of the mass gap is required. We would also like to
obtain a better understanding of the full spectrum and explicit
expressions for the higher--order eigenvectors in terms of
three--dimensional partitions. Based on the number of conserved
quantities, one is inclined to believe also the 3d quantum system to be
integrable. If this is indeed the case is another a question worth
studying.  In this vein one could also try to generalize the Bethe
ansatz for the \textsc{xxz} spin chain to the constrained system of
spin chains corresponding to the 3d case.  Furthermore, we wonder
whether the $SU_q(2)$ quantum symmetry of the \textsc{xxz} spin chain
carries over to the 3d melting quantum crystal.


%% file: NPointFunctions.tex
\section{One--point function}
\label{sec:calculations}

In this appendix we describe in detail the computation for the
one--point function for the two-dimensional kink introduced in
Sec.~\ref{sec:one-point-function}.

The starting point is the one--point function for the generating
function of the kinks. Given the representation in terms of direct
product over the spins (grand canonical ensemble) one finds:
\begin{equation}
  \braket{\Psi(z) | \sigma_x^3 | \Psi(z)} = \frac{1 - w\, q^{x+1/2}}{2\left(1 + w \,q^{x+1/2}\right)}   ( -w\, q^{1/2} ;\, q )_\infty \, .
\end{equation}
Using the development
\begin{equation}
  \frac{1 - w \zeta}{1 + w \zeta} = -1 + 2 \sum_{k=0}^\infty \left( -1 \right)^k w^k \zeta^k \, ,
\end{equation}
one can expand:
\begin{equation}
  \braket{\Psi(z) | \sigma_x^3 | \Psi(z)} = - \frac{1}{2}  ( -w\, q^{1/2} ; q )_\infty  + \sum_{N=0}^\infty \sum_{k=0}^N \left( -1 \right)^{N-k} q^{\left( x+1/2\right) \left( N - k \right)} \frac{q^{k^2/2}}{(q;q)_k} w^N \, .
\end{equation}
The contribution of the $N$--particle kink is therefore
\begin{equation}
  m_N (x) = \frac{\braket{\Psi_N | \sigma^3 | \Psi_N}}{\braket{\Psi_N|\Psi_N}} = -\frac{1}{2} + \sum_{k=0}^{N} \qbinom{N}{k}_{q} \left( -1  \right)^k q^{\left( x + 1/2 \right) k} q^{-k (N - k/2)} (q;q)_k \, ,  
\end{equation}
where $ \qbinom{N}{k}_{q}$ is the $q$--binomial coefficient
\begin{equation}
  \qbinom{N}{k}_q = \frac{(q;q)_N}{(q;q)_{N-k} (q;q)_k} \, .  
\end{equation}
Using the identity (see~\cite{Gasper:1990bh})
\begin{equation}
  \qbinom{N}{k}_{q} = \frac{ ( q^{-N}; q)_k}{(q;q)_k} \left( - 1 \right)^k q^{N k} q^{- \binom{k}{2}} \, ,
\end{equation}
one finds
\begin{equation}
  m_N (x) = -\frac{1}{2} +  \sum_{k=0}^N (q^{-N} ; q)_k q^{k \left( x + 1 \right)} \, .  
\end{equation}
This represents a kink centered in $x=N$. For later convenience it is preferable to shift the center to $x=0$:
\begin{equation}
  m_N (x) = -\frac{1}{2} + \sum_{k=0}^N (q^{-N} ; q)_k q^{N k} q^{k x} \, .
\end{equation}
In the $N \to \infty$ limit one can use
\begin{equation}
  \lim_{N \to \infty} (q^{-N}; q)_k q^{N k} = \left( - 1 \right)^k q^{\binom{k}{2}} \, ,  
\end{equation}
and find the \emph{exact} expression for the shape of the kink
\begin{equation}
  m_\infty(x) = -\frac{1}{2} +  \sum_{k=0}^\infty \left( - 1 \right)^k q^{\binom{k}{2}} q^{k x} \, .  
\end{equation}
To study the $q \to 1$ limit, introduce the function
\begin{equation}
  f_q (\zeta) = \sum_{k = 0} \left( -1  \right)^k  q^{\binom{k}{2}} \zeta^k \, .
\end{equation}
One can verify that $f_q(\zeta)$ satisfies the $q$--difference equation
\begin{equation}
  \zeta \, f_q (q\,\zeta) = 1 - f_q(\zeta) \, ,
\end{equation}
and introducing the $q$--difference operator as in~\cite{Gasper:1990bh}:
\begin{equation}
  D_q f(\zeta) = \frac{f(\zeta) - f(q\,\zeta)}{\left(1 - q \right)\zeta} \, ,  
\end{equation}
one obtains
\begin{equation}
  \left( 1 - q \right) D_q f_q(\zeta) = - \frac{1}{\zeta^2} + \frac{1+\zeta}{\zeta^2} f_q(\zeta) \, .
\end{equation}
In the $q \to 1^-$ limit, $D_q$ becomes the derivative with respect to
$\zeta$. Define
\begin{equation}
\label{eq:limit-fq}
  f(x) = \lim_{q \to 1^-} f_q (q^{-x/ \log(q)}) \, .  
\end{equation}
Solving the differential equation for $f(x)$, one obtains
\begin{equation}
  f(x) = \lim_{q \to 1^-} \frac{1}{q-1} \exp [ \tfrac{e^x}{q-1} + x] E_{\nicefrac{1}{\left(q - 1 \right)}} (\tfrac{e^x}{q-1})  \, ,
\end{equation}
where $E_n (x) $ is the exponential integral
\begin{equation}
  E_n (x) = \int_1^\infty \frac{e^{-xt}}{t^n} \di t \, .  
\end{equation}
Using the asymptotic expansion for $n\to \infty$:
\begin{equation}
  E_n (x) \sim \frac{e^{-x}}{x} \frac{1}{1+n/x} \, ,  
\end{equation}
we obtain the asymptotic expression for $f(x)$:
\begin{equation}
  f(x) = \frac{e^x}{1+e^x} \, .  
\end{equation}
It follows that the magnetization profile is
\begin{equation}
  m (x) = -\frac{1}{2} + f(x) = -\frac{1}{2} \tanh (\frac{x}{2}) \, .  
\end{equation}
The shape of the partition is therefore given by
\begin{equation}
  \mu_\phi (x) = - 2\int \di x \; m_\phi (x) = - 2 \log ( 2\,\cosh( \tfrac{x}{2} ))
\end{equation}
or, in a more symmetric form (take $\mu_\phi (x)=:y$),
\begin{equation}
  e^{(x-y)/2} + e^{-(x+y)/2} = 1 \, .  
\end{equation}
